\documentstyle[aps,epsfig]{revtex}

\newcommand{\be}{\begin{equation}}
\newcommand{\ee}{\end{equation}}

\def\la{\mathrel{\mathpalette\fun <}}

\def\fun#1#2{\lower3.6pt\vbox{\baselineskip0pt\lineskip.9pt
\ialign{$\mathsurround=0pt#1\hfil##\hfil$\crcr#2\crcr\sim\crcr}}}

\begin{document}

\title{Decay $\phi (1020)\to \gamma f_0(980)$:
analys in the nonrelativistic quark model approach }
\author{A. V. Anisovich, V. V. Anisovich, V. N. Markov,\\ V. A.
Nikonov, A. V. Sarantsev}
\date{30.12.2004}
\maketitle

\begin{abstract}

We demonstrate the possibility of a good description of the processes
$\phi (1020)\to \gamma\pi\pi$
and $\phi(1020)\to \gamma f_0(980)$ within the
framework of nonrelativistic quark model assuming $f_0(980)$ to be
dominantly quark--antiquark system. Different mechanisms of the
radiative decay, that is, the emission of photon by the
constituent quark (additive quark model)
and charge-exchange current, are considered.
We also discuss the status of the threshold theorem
applied to the studied reactions, namely, the
behavior of the decay amplitude at $M_{\pi\pi} \to m_\phi$ and
$m_{f_0}\to m_\phi$. In conclusion the  arguments favoring the
$q\bar q$ origin of $f_0(980)$ are listed.

\end{abstract}

\section{Introduction}

$\;$ The $K$-matrix analysis of meson spectra  [1--3]
%\cite{kmat,prokoshkin,km96}
and meson systematics \cite{syst,ufn} point
determinatedly to the quark--antiquark origin of $f_0(980)$. However
there exist hypotheses where $f_0(980)$ is interpreted as four-quark
state \cite{jaffe}, $K\bar K$ molecule \cite{isgur} or vacuum scalar
\cite{close-v}. The radiative and weak decays involving $f_0(980)$ may
be decisive tool for  understanding  the nature of $f_0(980)$.

In the present paper the reaction $\phi(1020)\to \gamma f_0(980)$ is
considered in terms of nonrelativistic quark model assuming
$f_0(980)$ to be dominantly the $q\bar q$ state.
 nonrelativistic quark model is a good approach for the description
 of the lowest $q\bar q$ states of pseudoscalar and vector nonets, so
one may hope that the lowest scalar $q\bar q$ states are described
with reasonable accuracy as well.
The choice of nonrelativistic approach for the analysis of the
reaction $\phi (1020)\to \gamma f_0(980)$ was motivated by the fact that
in its framework we can take account of not only the additive quark
model processes (emission of the photon by constituent quark) but also
those beyond it within the use of the dipole formula (the photon
emission by the charge-exchange current gives such
an example). The dipole formula
for the radiative transition of vector state to scalar
one, $V\to \gamma S$ was applied before for the calculation of
reactions with heavy quarks, see \cite{8,9}
and references therein. Still, a straightforward
application of the dipole formula to the reaction $\phi (1020)\to
\gamma f_0(980)$ is hardly possible, for the $f_0(980)$ for sure cannot
be represented as a stable particle: this resonance is characterized by
two poles laying on two different sheets of the comlex-$M$ plane , at
$M=1020-i40$ MeV and $M=960-i200$ MeV. It should be emphasized that
these two poles are important for the description of $f_0(980)$.
Therefore we use below the method as follows: we calculate the
radiative transition to a stable bare $f_0$ state (this is
$f_0^{\rm bare}(700\pm 100)$, its parameters were obtained in the
$K$-matrix analysis \cite{kmat}. In
 this way we find out the description of the process
$\phi(1020)\to \gamma f_0^{\rm bare}(700\pm 100)$ and furthermore we switch
on the hadronic decays and determine the transition
$\phi(1020)\to \gamma\pi\pi$; just the residue in the pole of this
amplitude is  the radiative transition amplitude
$\phi(1020)\to \gamma f_0(980)$. Hence we obtain a successful
description of data for $\phi(1020)\to \gamma\pi\pi$   and
$\phi(1020)\to \gamma f_0(980)$ within the assumption that  $f_0(980)$
is dominated by the quark--antiquark state.

The conclusion about the nature of $f_0(980)$ cannot be based on
the study of one
reaction only but should be motivated by the whole aggregate  of data.
In the article, we list also the other processes, which provide us
with arguments in favor of the dominant $q\bar q$ structure of
$f_0(980)$.

Section 2 is introductive: here we consider a simple model
for the description of
composite vector ($V$) and scalar ($S$) particles, the composite
particles consisting of one-flavor quark,  charge-exchange
currents being absent. In such a model the decay transition
$V\to \gamma S$ is completely determined by the additive quark model
process: the photon is emitted only by
one or another constituent quark. Two
alternative representations of the $V\to \gamma S$ decay  amplitude are
given, namely,  the standard additive quark model
formula and that of a photon
dipole emission, in the latter the factor $\omega=(m_V-m_S)$ is
written in the explicit form. The comparison of these two
representations helps us to formulate the problem of application of the
threshold theorem \cite{siegert} to the reaction $V\to \gamma S$.
Using a simple example with exponetial wave functions, we
demonstrate the $\omega^3$ factor occurred in the partial
decay width when the transition $V\to \gamma S$ is considered in terms
of the additive quark model.

The threshold theorem has a straightforward formulation
for the stable $V$ and $S$ states but it is not the case for
resonances, which are the main objects of our present study. That is
why we intend to reformulate the threshold theorem as the requirement
of the amplitude analyticity -- this is given in Section 3, on the
basis of \cite{gauge}. Working with  nonstable particles, when $V$
and $S$ are resonances, the $V\to \gamma S$ amplitude
should be determined as a  residue of a more general
amplitude, with stable particles in the initial and final states. For
example, the $\phi (1020)\to \gamma f_0 (980)$ amplitude should be
defined as a residue of the $e^+e^- \to \gamma\pi\pi$ amplitude in the
poles corresponding to resonances $\phi (1020)$ and $f_0 (980)$ (the
$\phi$ pole in the $e^+e^-$-channel and the $f_0$ pole in the $\pi\pi$
channel). The only residue in the pole defines the universal amplitude
which does not depend on the considered reaction.

In Section 4 we discuss the reaction $\phi\to \gamma f_0$ for
the case, when the $f_0$ is a multicomponent system and  $f_0$ and
$\phi$ are stable
states with respect to hadronic decays. The analysis of
meson spectra (e.g. see the latest $K$-matrix analyses
\cite{kmat,ufn}) definitly tells us that the $f_0$-mesons are the
mixture of the quarkonium ($n\bar n=(u \bar u +d\bar d)/\sqrt{2}$ and
$s\bar s$)  and gluonium components. Such a multichannel structure
of $f_0$ states reveals itself in the existence of the $t$-channel
charge-exchange currents.
Therefore, the transition $\phi\to \gamma f_0$ goes via two mechanisms:
the photon emission by  constituent quarks (additive quark
model process) and charge-exchange current.
We write down the formulae
for the amplitudes  initiated by these two mechanisms. The equality to
zero of the whole amplitude at small mass difference of $\phi$ and
$f_0$, $A_{\phi\to \gamma f_0}\sim \omega$ at $\omega\to 0$, is
resulted from the cancellation of
contributions of these two mechanisms. We also give
a dipole representation of the transition amplitude, where $A_{\phi\to
\gamma f_0}$ is determined through the mean transition radius and mass
difference $(m_\phi-m_{f_0})$.

Section 5 is devoted to the reaction $\phi(1020)\to \gamma f_0(980)$.
 First,  we discuss whether it is possible to treat $f_0(980)$
as a stable particle. Our answer is "no",
in fact the $f_0(980)$ is
unstable particle  characterized by two poles
located near the $K\bar K$ threshold.  As  was stressed above, strong
transitions $f_0(980) \to \pi\pi,\, K\bar K$
reveal themselves in the two amplitude poles, which
are located on different sheets of the complex-$M$ plane, at
$M=1020-i\,40$ MeV   and $M=960-i\,200 $ MeV, and
these two poles are important for the description of $f_0(980)$. The
essential role of the second pole is seen by considering  the $\pi\pi$
spectrum in $\phi (1020)\to \gamma\pi\pi$ (Section 6): the
visible width of the
pick in the $\pi\pi$  spectrum is of the order of $150$ MeV,
and the spectrum decreases slowly with further decrease of $M_{\pi\pi}$.

Another problem to be discussed is the choice of a method for the
consideration of radiative decay amplitude. One can work within two
alternative representations for the $A_{\phi\to \gamma f_0}$ amplitude.
One representation uses additive quark model
complemented with the contribution from the
charge-exchange current processes.
Whence the additive quark model amplitude can
be calculated rather definetely, at least for the lowest $q\bar q$
states, the charge-exchange current processes are vaguely determined.

The other way to deal with the transition amplitude consists in using
the dipole emission formulae, where the amplitude is defined  by the
mean transition radius  and factor $(m_\phi-m_{f_0})$. For the lowest
$q\bar q$ states, we have a good estimate of the radius.
But there is a problem of the determination of the factor
$(m_\phi-m_{f_0})$, because the $f_0(980)$
is unstable particle characterized by two
poles. The pole $M=960-i\,200$ MeV is disposed on the same sheet of the
complex-$M$ plane as the pole of $\phi$-meson, and the distance
$|m_\phi-m_{f_0}|$ is $\sim$200 MeV, while the pole $M=1020-i\,40$ MeV
is located on another sheet, the distance from the $\phi$ meson is
$\sim$70 MeV, that is also not small in the hadronic scale. The problem
is what the mass difference factor means in case of complex masses
and which pole should be used to characterise this mass difference.

To succeed in the description of the decay
$\phi(1020)\to \gamma f_0(980)$
we use the results of the $K$-matrix analysis of
the $(IJ^{PC}=00^{++})$ wave \cite{kmat}. The fact is that, on the one
hand, the $K$-matrix analysis allows us to get the experimentally based
information on masses and full widths of resonances
together with the pole residues needed for the decay couplings and
partial widths. On the other hand, the knowledge of the $K$-matrix
amplitude enables us to trace the evolution of states by switching
on/off the decay channels. In such a way, one may obtain the
characteristics of the bare states, which are predecessors of real
resonances. With such characteristics, one can perform a reverse
procedure: to retrace the transformation of the amplitude written in
terms of bare states to the amplitude corresponding to the trasition to
real resonance. Just this procedure has been applied in Section 5 for
the calculation of the decay amplitude $\phi(1020)\to \gamma f_0(980)$.

Therefore, within the frame of
nonrelativistic quark model, we have calculated the
reaction $\phi(1020)\to \gamma f_0^{\rm bare}(n)$,
where $f_0^{\rm bare}(n)$ are bare states found in \cite{kmat}.
 Furtheremore, with the
$K$-matrix technique, we have taken account of the decays
$f_0^{\rm bare}(n)\to \pi\pi, KK$,
thus having calculated the reaction $\phi (1020)\to \gamma\pi\pi$
and the amplitude of $\phi(1020)\to \gamma f_0(980)$ (the pole
residue in the $\pi\pi$ channel). In this way
we see that the main contribution is
given by the transition $\phi(1020)\to \gamma f_0^{\rm bare}(700\pm100)$.
The characteristics of $f_0^{\rm bare}(700\pm 100)$ are fixed by the
$K$-matrix analysis \cite{kmat}:
this is a $q\bar q$ state close to the flavor
octet and it is just the predecessor of $f_0(980)$. In the
framework of this approach we succeed in the description of data for
the reactions
$\phi(1020)\to \gamma f_0(980)$ (Section 5) and
$\phi (1020)\to \gamma\pi\pi$ (Section 6).

Let us note that such a method, the use of bare states for
the calculation of meson spectra, has been applied before for the study
weak hadronic decays $D^+\to \pi^+\pi^+\pi^-$ \cite{ital} and
description of the $\pi\pi$ spectra in photon--photon
collisions $\gamma\gamma\to \pi\pi$ \cite{ale-K}.

The question of what is the accuracy of the additive quark model in the
description of the reactions
$\phi(1020)\to \gamma f_0^{\rm bare}(700\pm100)$ and
$\phi(1020)\to \gamma f_0(980)$ is discussed in Section 7. We
compare the results of the calculation of the
$\phi(1020)\to \gamma f_0^{\rm bare}(700\pm100)$ reaction by using the dipole
formula with that of the additive quark model. It is seen that, within
error-bars given by the $K$-matrix analysis \cite{kmat}, the results
coincide. Still, one should emphasize that the dipole-calculation
accuracy is low that is due to a large error in the determination of
the bare-state masses.
 The coincidence of results in the dipole and additive
quark model formulae should point to a small contribution of processes
which violate additivity such as  photon emission by the charge-exchange
current: this smallness is natural, provided the hadrons are
characterized by two sizes, namely, the hadron radius $(R_h\sim
R_{\rm conf})$ and constituent-quark radius $(r_q)$ under the
condition $r_q^2\ll R_h^2$, see \cite{book} and references therein.

The performed analysis demonstrates that the studied reaction,
$\phi(1020)\to \gamma f_0(980)$ does not provide us any difficulty
with the interpretation of $f_0(980)$ as  $q\bar q$-state. Still, to
conclude about the content of $f(980)$ we list
 in Section 8 the arguments in favor
of the $q\bar q$ origin of $f_0(980)$.

\section{The process $V\to\gamma S$ within nonrelativistic additive
quark model}

$\;$Here, in
 the framework of  nonrelativistic quark model, we consider
 the  transition $V\to \gamma S$
in  case when the charge-current exchange forces are absent
and the $V\to \gamma S$ amplitude is given by the additive quark model
contribution.

\subsection{Wave functions for  vector and scalar composite
particles}

$\;$ The $q\bar q$ wave functions of vector $(V)$ and scalar $(S)$
particles are defined as follows:
\begin{eqnarray}
\Psi_{V\mu}({\bf k}) &=& \sigma_\mu\psi_V(k^2), \nonumber\\
\Psi_S({\bf k}) &=& (\mbox{\boldmath$\sigma \cdot k$})\psi_S(k^2),
\label{1}
\end{eqnarray}
where, by using Pauli matrices, the spin factors are singled out.
The blocks dependent on the relative momentum squared
are related to the vertices in the following way:
\begin{eqnarray}
\psi_V(k^2) &=& \frac{\sqrt m}2\ \frac{G_V(k^2)}{k^2+m\varepsilon_V},
\nonumber\\
\psi_S(k^2) &=& \frac1{2\sqrt m}\ \frac{G_S(k^2)}{k^2+m\varepsilon_V}\
.
\label{2}
\end{eqnarray}
Here $m$ is the quark mass,  $\varepsilon$ is the
composite-system binding energy:
 $\varepsilon_V=2m-m_V$ and $\varepsilon_S=2m-m_S$, where
$m_V$ and $m_S$ are the masses of bound states.
The
normalization condition for the wave functions reads
\begin{eqnarray}
&& \int\frac{d^3k}{(2\pi)^3}\mbox{ Sp}_2\left[\Psi^+_S ({\bf k})
\Psi_S({\bf k})\right]=\int\frac{d^3k}{(2\pi)^3}\,\psi^2_S(k^2)
\mbox{ Sp}_2[(\mbox{\boldmath$\sigma \cdot k$})
(\mbox{\boldmath$\sigma \cdot k$})]
=1, \\
&& \int\frac{d^3k}{(2\pi)^3}\mbox{ Sp}_2\left[\Psi^+_{V\mu} ({\bf
k}) \Psi_{V\mu'}({\bf k})\right]=\int\frac{d^3k}{(2\pi)^3}\,
\psi^2_V(k^2) \mbox{ Sp}_2[\sigma_\mu\sigma_{\mu'}]=\delta_{\mu\mu'} .
\nonumber
\label{3}
\end{eqnarray}

\subsection{Amplitude within additive quark model}

 $\;$ When a photon is emitted by quark or antiquark,
the $V\to \gamma S$ process is described by the triangle diagram,
see Fig. 1$a$, that is actually the contribution from  additive quark
model.
Relativistic consideration of the triangle diagram
is presented in \cite{epja,s}, while the discussion of
nonrelativistic approximation is given in \cite{gauge,nonrel}
(recall that in \cite{nonrel} corresponding wave functions were
determined in another way, namely:
$\psi_V(k^2)=G_V(k^2)(4k^2+4m\varepsilon_V)^{-1}$ and
$\psi_S(k^2)=G_S(k^2) (4k^2+4m\varepsilon_S)^{-1}$).

In terms of  wave functions
(\ref{1}) the triangle-diagram contribution reads
\be
\label{4}
\epsilon^{(V)}_\mu\epsilon^{(\gamma)}_\alpha
A^{V\to\gamma S}_{\mu\alpha}
=eZ_{V\to \gamma S}              \,
\epsilon^{(V)}_\mu\epsilon^{(\gamma)}_\alpha
F^{V\to\gamma S}_{\mu\alpha} \, ,
\ee
$$
F^{V\to\gamma S}_{\mu\alpha}
= \int
\frac{d^3k}{(2\pi)^3}\mbox{ Sp}_2\left[\Psi^{+}_S({\bf
k})4k_\alpha\Psi_{V\mu}({\bf k})\right]\, .
$$
Here
$\epsilon^{(V)}_\mu$ and $\epsilon^{(\gamma)}_\alpha$
are polarization vectors for $V$ and $\gamma$:
$\epsilon^{(V)}_\mu p_{V\mu} =0$ and $\epsilon^{(\gamma)}_\alpha
q_\alpha =0$.
The charge factor
$Z_{V\to \gamma S}$
being different for different reactions is specified below (see
also \cite{epja,s}).
The expression for the transition amplitude (\ref{4})
can be simplified after the
 substitution in the integrand
\be
\mbox{Sp}_2[\sigma_\mu(\mbox{\boldmath$\sigma \cdot k$})]\ k_\alpha\
\rightarrow\ \frac23\ k^2\,g^{\perp\perp}_{\mu\alpha},
\label{5}
\ee
where $g^{\perp\perp}_{\mu\alpha}$ is the metric tensor
in the space orthogonal to total
momentum of the vector particle $p_V$ and
photon $q$. The substition (\ref{5}) results in:
\be
\label{5a}
A^{V\to\gamma S}_{\mu\alpha}
=eg^{\perp\perp}_{\mu\alpha}
A_{V\to\gamma S}\, ,
\ee
where
\be
\label{5b}
A_{V\to\gamma S}
=Z_{V\to \gamma S}
\int \limits_0^\infty
\frac{dk^2}{\pi}\psi_S(k^2)\psi_V( k^2)\frac {2}{3\pi}k^3  .
\ee
The amplitudes
$A^{V\to\gamma S}_{\mu\alpha}$ and $A_{V\to\gamma S}$ were
used in \cite{epja,s} for
the decay amplitude $\phi(1020)\to \gamma f_0(980)$
within  relativistic treatment of the quark
transitions.

 However, for our purpose it would be suitable  not to deal
with Eq. (\ref{5b}) but  use the form factor
$F^{V\to\gamma S}_{\mu\alpha}$  of
Eq. (\ref{4}) rewritten in the
coordinate representation. One has:
\begin{eqnarray}
&& \Psi_{V\mu}({\bf k})\ =\int d^3r\ e^{i\bf k\cdot r}\
\Psi_{V\mu}({\bf r}), \nonumber\\
&& \Psi_S({\bf k})\ =\int d^3r\ e^{i{\bf k\cdot r}}\
\Psi_S({\bf r}).
\label{6}
\end{eqnarray}
Then the form factor
$F^{V\to\gamma S}_{\mu\alpha}$
 can be represented as follows:
\be
F^{V\to\gamma S}_{\mu\alpha}=\
 \int d^3r\mbox{
Sp}_2\left[\Psi^{+}_S({\bf r}) 4k_\alpha\Psi_{V\mu}({\bf
r})\right],
\label{7}
\ee
where $k_\alpha$ is the operator: $k_\alpha=-i\nabla_\alpha$.
This operator can be written as the commutator of $r_\alpha$ and
$-\nabla^2/m=T$ (kinetic energy):
\be
2 i\;m(Tr_\alpha-r_\alpha T)\ =\ 4(-i\nabla_\alpha).
\label{8}
\ee

Let us
consider the case when the quark--quark interaction is rather simple,
say, it depends on the relative interquark distance with the
potential $U(r)$. For vector and scalar composite systems
we also use  additional  simplifying assumption:
vector and scalar mesons consist of quarks of the same flavor $(q\bar
q)$. Then we have the following Hamiltonian:
\be
H\ =\ -\ \frac{\nabla^2}m+U(r),
\label{9}
\ee
and can rewrite (\ref{8}) as
\be
2i\;m(Hr_\alpha-r_\alpha H)\, =\ 4(-i\nabla_\alpha).
\label{10}
\ee
After substituting the commutator in (\ref{7}),
the transition form factor for the reaction $V\to\gamma
S$ reads
\be
 F^{V\to\gamma S}_{\mu\alpha}\, =\,\int d^3r\mbox{
Sp}_2\left[\Psi^{+}_S(r)r_\alpha\Psi_{V\mu}(r)\right]2i\;m
(\varepsilon_V-\varepsilon_S).
\label{11}
\ee
Here we have used that  $(H+\varepsilon_V)\Psi_V=0$  and
$(H+\varepsilon_S)\Psi_S=0$.

The factor $\varepsilon _V-\varepsilon_S$
in the right-hand side (\ref{11}) is a manifestation of the
threshold theorem: at
$\varepsilon_V-\varepsilon_S )=m_S-m_V\to0$ the form factor
$F^{V\to\gamma S}_{\mu\alpha}$ turns to zero.
Actually, in the additive quark model the amplitude
of the $V\to\gamma S $ transition,
being determined by the process of Fig. 1$a$,
cannot be zero if $V$ and $S$
are basic states with radial quantum number $n=1$:
in this case the wave functions
$\psi_V(k^2)$ and $\psi_S(k^2)$ do not change
sign, and the right-hand side (\ref{5b}) does not equal
zero.  In order to clarify this point let us consider as an example the
exponetial approximation for the wave functions $\psi_V(k^2)$ and
$\psi_S(k^2)$.

\subsection{Basic vector and scalar $q\bar q$ states: the example of
exponential approach to wave functions}

$\;$ We parametrize the ground-state wave functions of scalar
and vector particles as follows:
\begin{eqnarray}
\Psi_{V_\mu}(r)=\sigma_\mu\psi_V(r^2), &&
\psi_V(r^2)=\frac1{2^{5/4}\pi^{3/4}b^{3/4}_V}\exp\left[-\frac{r^2}{4b_V}
\right], \nonumber\\
&& \label{11a}\\
\Psi_S(r)=(\mbox{\boldmath$\sigma \cdot r$})\psi_S(r^2), &&
\psi_S(r^2)=\frac i{ 2^{5/4}\pi^{3/4}b_S^{5/4}\sqrt3}
\exp\left[-\frac{r^2}{4b_S}\right].
\nonumber
\end{eqnarray}

The wave functions with $n=1$ have no nodes;
numerical factors take account of the normalization conditions
\begin{eqnarray}
 \int d^3r\mbox{ Sp}_2\left[\Psi^+_{S}(r)\Psi_{S}
(r)\right]=\ 1\, ,
&& \int d^3r\mbox{ Sp}_2\left[\Psi^+_{V\mu}(r)\Psi_{V\mu'}
(r)\right]=\ \delta_{\mu\mu'}.
\label{11b}
\end{eqnarray}

With exponential wave functions, the matrix element for
$V\to\gamma S$ given by the additive quark model diagram, Eq. (\ref{7}),
is equal to
\be
\label{11c}
\epsilon^{(V)}_\mu\epsilon^{(\gamma)}_\alpha
F^{V\to\gamma S}_{\mu\alpha}({\rm additive})=
(\epsilon^{(V)}\epsilon^{(\gamma)})\,
\frac{2^{7/2}}{\sqrt3} \frac{b^{3/4}_Vb^{5/4}_S}{(b_V+b_S)^{5/2}}.
\ee

Formula for $F^{V\to\gamma S}_{\mu\alpha}$
written in the frame of the dipole
emission, Eq. (\ref{11}), reads
\be
\label{11d}
\epsilon^{(V)}_\mu\epsilon^{(\gamma)}_\alpha
F^{V\to\gamma S}_{\mu\alpha}({\rm dipole})=
(\epsilon^{(V)}\epsilon^{(\gamma)})
\frac{2^{7/2}}{\sqrt3}\,\frac{b^{7/4}_Vb^{5/4}_S}{(b_V+b_S)^{5/2}}
\, m(m_V-m_S).
\ee

In case under consideration
(one-flavor quarks with Hamiltonian given by Eq. (\ref{9})), the
equations (\ref{11c}) and (\ref{11d}) coincide,
$F^{V\to\gamma S}_{\mu\alpha}{\rm (additive)}=
F^{V\to\gamma S}_{\mu\alpha}{\rm (dipole)}$,
therefore
\be
\label{11e}
m(m_V-m_S)\ =\ b^{-1}_V,
\ee
that means
that the factor $\varepsilon_S-\varepsilon_V$ in the right-hand
side (\ref{11}) relates to the difference between the
$V$ and $S$ levels and is defined by  $b_V$ only.  In this
way, the form factor
$F^{V\to\gamma S}_{\mu\alpha}$ turns to zero only when $b_V $
(or $b_S$) tends to the infinity.

The considered example does not mean that the threshold theorem
for the reaction
$V\to\gamma S$ does not work,  this tells us only that we should
interprete and use it carefully.
In the next Section, we discuss how to formulate the threshold
theorem based on the requirement of amplitude analyticity, thus getting
more information on the threshold theorem applicability.

\section{Analyticity of the amplitude and the threshold
theorem}

$\;$ The threshold theorem can be formulated as the requirement of
analyticity of the amplitude. To clarify this statement we
consider here not only the transition of the bound states
but more
general process shown in Fig. 1$b$, where the interacting constituents
being in the vector $J^P=1^-$ state emit the photon and then turn
into the scalar $J^P=0^+$ state.
This amplitude has as a subprocess the
bound state transition. Namely,
the blocks for the rescattering of
constituents in Fig. 1$b$ contain the poles related to bound states,
see Fig. 1$c$, and the residues in these poles determine
the bound-state transition amplitude (triangle
diagram shown as intermediate block in Fig. 1$c$).

With the notations for invariant mass squares in the initial and final
  states of Fig. 1$b$ as follows
\be
P^2_V\ =\ s_V, \qquad P^2_S\ =\ s_S,
\label{12}
\ee
we can write the spin structures for this more general transition
 $V\to\gamma S$. The standard representation of this amplitude is
\be
A_{\mu\alpha}^{(V\to\gamma S)}(s_V,s_S,q^2\to 0)\, = \,
\left(g_{\mu\alpha}-\frac{2q_\mu P_{V\alpha}}{s_V-s_S}\right)
A_{V\to\gamma S}(s_V,s_S,0).
\label{13}
\ee
Here we stress
that the amplitude $A_{V\to\gamma S}$ describes the
emission of  real photon,
$q^2=0$. In (\ref{13}), it was taken into account that $(P_Vq)=
(s_V-s_S)/2$. The requirement of analyticity, i.e. the absence of
a pole at $s_V=s_S$, leads to the condition
\be
\bigg[A_{V\to\gamma S}(s_V,s_S,0)\bigg]_{s_V\to s_S}\ \rightarrow\ 0,
\label{14}
\ee
that is the threshold theorem for the transition amplitude
$V\to\gamma S$.

It should be now emphasized that the form of the spin factor in Eq.
(\ref{13}) is not unique. Alternatively, one can write the spin factor
as the metric tensor
$g^{\perp\perp}_{\mu\alpha}$ working in the space orthogonal to
$P_V$ and $q$, i.e.
$P_{V\mu}g^{\perp\perp}_{\mu\alpha}=0$ and
$g^{\perp\perp}_{\mu\alpha}\,q_\alpha=0$,
see Eq. (\ref{5}). This metric tensor reads
\be
g^{\perp\perp}_{\mu\alpha}(0)=g_{\mu\alpha}+\frac{4s_V}{(s_V-s_S)^2}\
q_\mu q_\alpha-\frac2{s_V-s_S}(P_{V\mu}q_\alpha+q_\mu P_{V\alpha}),
\label{15}
\ee
and we have used it in Eq. (\ref{5a}).
The uncertainty in the choice of  spin factor is due to the fact that
the difference
\be
\label{16}
g^{\perp\perp}_{\mu\alpha}(0)-\left(g_{\mu\alpha}- \frac{2q_\mu
P_{V\alpha}}{s_V-s_S}\right)=\ 4L_{\mu\alpha}(0),
\ee
where
\be
\label{17}
L_{\mu\alpha}(0)\ =\ \frac{s_V}{(s_V-s_S)^2}\,q_\mu q_\alpha-
\frac1{2(s_V-s_S)}\,P_{V\mu}q_\alpha,
\ee
is the nilpotent operator  \cite{gauge}
\be
L_{\mu\alpha}(0)L_{\mu\alpha}(0)\ =\ 0.
\label{18}
\ee
The addition of the nilpotent operator
$L_{\mu\alpha}(0)$ to  spin factor of the  transition amplitude
 $V\to\gamma S$ does not change the expression
 $A_{V\to\gamma S}(s_V,s_S,0)$, see \cite{gauge} for more detail.
Here, by discussing  the analytical structure of the amplitude,
it is convenient to
work with the operator (\ref{13}), for
it is the least cumbersome.

Consider now the reaction $V\to\gamma S$ ($V$ and $S$ being
quark--antiquark bound states), say, of the type of $\phi\to\gamma
f_0$ or $\phi\to\gamma a_0$.
Because of the confinement the quarks are not the particles which
form the $|{\rm in}\rangle$ and $\langle {\rm out}|$ states, therefore
the amplitudes like $A_{\phi\to\gamma f_0}$ are to be defined as
 the amplitude
 residue for
the process with the scattering of the stable particles, for
example, for $e^+e^-\to\gamma\pi^+\pi^-$ (see Fig. 2):
\begin{eqnarray} &&
A^{(e^+e^-\to\gamma\pi^+\pi^-)}_{\mu\alpha} (s_V,s_S,0)\ =\left(
g_{\mu\alpha}-\frac{2q_\mu P_{V\alpha}}{s_V-s_S}\right)\times
\\
 &&\times \left[G_{e^+e^-\to\phi}\
\frac{A_{\phi\to\gamma f_0}(m^2_\phi,m^2_{f_0},0)}{(s_V-m^2_\phi)
(s_S-m^2_{f_0})}\ g_{f_0\to\pi^+\pi^-} +B(s_V,s_S,0)\right].
\nonumber
\label{19}
\end{eqnarray}
We see that
$A(m^2_\phi,m^2_{f_0},0)$, up to the factors $G_{e^+e^-\to\phi}$ and
$g_{f_0\to\pi^+\pi^-}$, is the residue in the amplitude poles
 $s_V=m^2_\phi$ and
$s_S=m^2_{f_0}$: just this value supplies us with the transition
amplitude for  the reactions with
bound states $\phi\to\gamma f_0$. If we deal with stable
composite particles, in other words, if $\phi$ and $f_0$ can be
included into the set of fields
$|{\rm in}\rangle$ and $\langle {\rm out}|$,
the transition amplitude $\phi\to\gamma f_0$ can be written in the form
similar to
 (\ref{13}):
\be
\label{20}
A^{(\phi\to\gamma f_0)}_{\mu\alpha}(m^2_\phi,m^2_{f_0},0)\
=\left(g_{\mu\alpha} -\frac{2q_\mu
p_\alpha}{m^2_\phi-m^2_{f_0}}\right)
A_{\phi\to\gamma f_0}\left(m^2_\phi,m^2_{f_0},0\right),
\ee
where we have substituted
$P_V\to p$.
For $A_{\phi\to\gamma f_0}\left(m^2_\phi,m^2_{f_0},0\right)$
the threshold theorem is fulfilled:
\be
\label{21}
\left[A_{\phi\to\gamma f_0}(m^2_\phi,m^2_{f_0},0)\right]_{m^2_\phi
\to m^2_{f_0}}\ \sim\ (m^2_\phi-m^2_{f_0}),
\ee
that means that the threshold theorem of Eq. (\ref{21}) reveals itself
as a requirement of analyticity of the amplitude $\phi\to\gamma f_0$
determined by Eq. (\ref{20}).

Let us emphasize again that formula (\ref{20})  has been written for
the $\phi$ and $f_0$ mesons  assuming them to be stable, i.e. they can
be treated as the states which belong to the sets
$|{\rm in}\rangle$ and $\langle {\rm out}|$.
However, by considering the process $\phi \to \gamma
f_0$, we deal with resonances, not  stable particles, and whether this
assumption is valid for resonances is a question which deserves special
discussion. We shall come back to this point below, and so far let us
investigate how the requirement (\ref{21}) is realized in quantum
mechanics, when $\phi$ and  $f_0$ are stable particles.

\section{Quantum mechanics consideration of the\\
reaction $\phi\to\gamma f_0$ with $\phi$ and
$f_0$ being\\ stable particles}

$\;$ In Section 2, we have considered the model for the reaction $V\to
\gamma S$, when $V$ and $S$ are formed by quarks of the same
flavor (one-channel model for $V$ and $S$).
The one-channel approach for $\phi (1020)$ (the dominance
of  $s\bar s$ component) looks acceptable, though for $f_0$ mesons it is
definitely not so:  scalar--isoscalar states are the mulicomponent ones.

The existence of several components in the $f_0$-mesons
changes the situation
with the $\phi \to \gamma f_0$ decays. First, the mixing of different
components may result in close values of
 masses of the low-lying vector
and scalar mesons. Second, equations (\ref{7}) and (\ref{11})
for the $\phi \to \gamma f_0$ decay turn to be nonequivalent
because of the photon emission by the $t$-channel exchange currents.

Here we consider
in detail a simple  model for $\phi$ and $f_0$:
the $\phi$ meson is treated as $s\bar
s$-system, with no admixture of the
nonstrange quarkonium, $n\bar n=(u\bar u+d\bar d)/\sqrt2$, nor
gluonium $(gg)$, while
the  $f_0$ meson is a mixture of $s\bar s$
and $gg$.

This model can be  considered
as a guide for the study of the reaction
$\phi(1020)\to\gamma f_0(980)$.
Indeed, the $\phi(1020)$ is almost
pure $s\bar s$ state, the admixture of the $n\bar n$ component in
$\phi(1020)$ is small, $\le 5\%$, and it can be neglected in a rough
estimate of the $\phi(1020)\to\gamma f_0(980)$ decay.

The resonance $f_0(980)$
is a multicomponent state.
Analysis of the $(IJ^{PC}=00^{++})$ wave in the $K$-matrix fit to
the data for meson spectra $\pi\pi$, $K\bar K$, $\eta\eta$,
$\eta\eta'$, $\pi\pi\pi\pi$  gives the following constraints for
the $s\bar s$, $n\bar n$ and $gg$-components in $f_0(980)$
\cite{kmat,ufn}:
\begin{eqnarray}
&&
50\%\ \la\ W_{s\bar s}[f_0(980)]\ <\ 100\%, \\ && 0\ \la\
W_{n\bar n}[f_0(980)]\ <\ 50\%, \nonumber\\ && 0\ \la\
W_{gg}[f_0(980)]\ <\ 25\%. \nonumber
\label{22}
\end{eqnarray}
Also, the $f_0(980)$ may contain a long-range $K\bar K$ component,
on the level of 10--20\%.

The restrictions (\ref{22}) permit the variant, when the probability
for the  $n\bar n$ component is small, and $f_0(980)$  is a mixture
of $s\bar s$ and $gg$ only. Bearing this variant in mind,
we consider such two-component model for $\phi$ and $f_0$ supposing
these particles are stable in respect to hadronic decays.

It is not difficult to generalise our consideration for the
three-component $f_0$  state
 ($n\bar n$, $s\bar s$ and $gg$),
 corresponding formulae are given in this Section too.

\subsection{Two-component model ($s\bar s$, $gg$) for $f_0$ and
$\phi $ }

$\;$ Now let us discuss the
 model, where $f_0$ has two components only: strange quarkonium
($s\bar s$ in the $P$ wave) and gluonium  ($gg$ in the $S$ wave).
The spin structure of the $s\bar s$
wave function is written in Section 2:
it contains the factor
({\boldmath$\sigma \cdot r$})  in the coordinate representation.
For the $gg$ system we have $\delta_{ab}$ or, in terms of
polarization vectors, the convolution
({\boldmath$\epsilon$}$_1^{(g)}${\boldmath$\epsilon$}$_2^{(g)})$.
Here we consider a simple
interaction, when the potential does not depend on
spin variables,
--- in this case one may forget about the vector structure of $gg$
working as if the gluon component consists of spinless particles.
As concern $\phi $, it is considered as a pure $s\bar s$ state
in the $S$ wave,
with the wave function
spin factor $\sim \sigma_\mu$, see Section 2.
So, the wave functions of $f_0$ and
$\phi$ mesons are written as follows:
\be
\label{31}
\hat\Psi_{f_0}({\bf r})\
=\left(\begin{array}{c}
\Psi_{f_0(s\bar s)}({\bf r})\\
\Psi_{f_0(gg)}(r)
\end{array}\right)
=\left(\begin{array}{c}
(\mbox{\boldmath$\sigma \cdot r$})\psi_{f_0(s\bar s)}(r)\\
\psi_{f_0(gg)}(r)
\end{array}\right)    \, ,
\ee
$$
\hat\Psi_{\phi\mu}({\bf r})\
=\left(\begin{array}{c}
\Psi_{\phi(s\bar s)\mu}({\bf r})\\
\Psi_{\phi(gg)\mu}(r)
\end{array}\right)
=\left(\begin{array}{c}
\sigma_\mu\psi_{\phi(s\bar s)}(r)\\
0
\end{array}\right)\, .
$$
The normalization condition is given by
Eq. (\ref{11b}), with the obvious replacement:
$\Psi_{S} \to \hat\Psi_{f_0}$ and
$\Psi_{V\mu} \to \hat\Psi_{\phi\mu}$.

The Shr\"odinger equation for the two-component states,
$s\bar s$ and $gg$, reads
\be
\label{23}
\left|\begin{array}{ccc}
\frac{k^2}{m}+U_{s\bar s\to s\bar s}(r)& , &  U_{s\bar s\to gg}({\bf
r})\\
 U^+_{s\bar s\to gg}({\bf r})& ,
& \frac{ k^2}{m_g}+U_{gg\to gg}(r)
\end{array}\right| \cdot \left(\begin{array}{c} \Psi_{s\bar s}({\bf
r}) \\ \Psi_{gg}({\bf r}) \end{array} \right)=\ E \left (
\begin{array}{c} \Psi_{s\bar s}({\bf r})\\ \Psi_{gg}({\bf r})
\end{array} \right ).  \ee
Furthermore we denote the Hamiltonian in
the left-hand side  (\ref{23}) as $H_0$.

We put the $gg$ component in $\phi$ to be
zero. It means that the potential $ U_{s\bar s \to gg}({\bf r})$
satisfies the following constraints:
\be
\langle 0^+s\bar s|  U_{s\bar s \to gg}({\bf r})|0^+ gg\rangle \neq 0,
\qquad
\langle 1^-s\bar s| U_{s\bar s \to gg}({\bf r})|1^-gg\rangle =0.
\ee
These constraints do not look  surprising for mesons in the region
1.0--1.5 GeV because the scalar  glueball is located just in this
mass region, while the vector one has considerably higher mass,
$\sim$2.5 GeV \cite{lattice}.

The $t$-exchange diagrams shown in Fig. 3$a$,3$b$,3$c$    are
the example of interaction leading to the potentials
$U_{s\bar s\to s\bar s}(r)$, $U_{gg\to gg}(r)$ and $ U_{s\bar s\to
gg}({\bf r})$. The potential
$ U_{s\bar s \to gg}({\bf r})$ contains the $t$-channel charge
exchange.

\subsubsection{Dipole emission of the photon in $\phi \to \gamma f_0$
decay}

$\;$ The Hamiltonian for the
interaction of  electromagnetic field with two-component
composite systems (quarkonium and gluonium components)
is presented in Appendix A.

For the transition $V\to\gamma S$, keeping the terms proportional to the
charge $e$, we have the following operator for the dipole emission:
\be \label{29}
\hat d_\alpha\ =\ \left|\begin{array}{ccc}
2k_\alpha & , & ir_\alpha U_{s\bar s\to gg}({\bf r})
\\ -ir_\alpha  U^+_{s\bar s\to gg}({\bf r})& , & 0
\end{array} \right|.
\ee
The transition form factor is given by the formula similar to Eq.
(\ref{7}) for the one-channel case, it reads
\be
\label{30}
F^{\phi\to\gamma f_0}_{\mu\alpha}\ =\
\int
d^3r\mbox{ Sp}_2\left[\hat\Psi^+_{f_0}({\bf r})\,2\hat d_\alpha
\hat\Psi_{\phi\mu}({\bf r})\right].
\ee
Drawing explicitly the two-component  wave functions,
one can rewrite Eq. (\ref{30}) as follows:
$$
F^{\phi\to\gamma f_0}_{\mu\alpha}\ =\
\int
d^3r\mbox{ Sp}_2\left[\Psi^+_{f_0(s\bar s)}({\bf r})\, 4
k_\alpha \Psi_{\phi (s\bar s)\mu}({\bf r})\right]
+
$$
\be
\label{30a}
+
\int
d^3r\mbox{ Sp}_2\left[\Psi^+_{f_0(gg)}({\bf r})\,
\left ( -ir_\alpha  U_{ gg\to s\bar s}({\bf r})\right )
\Psi_{\phi(s\bar s)\mu}({\bf r})\right].
\ee
The first term in the right-hand side (\ref{30a}), with the operator
$4k_\alpha$, is responsible for the interaction of a photon with
constituent quark that is the additive quark model contribution,
while the term $(-ir_\alpha U_{gg\to
s\bar s}({\bf r}))$ describes  interaction of the photon with the
charge flowing through the $t$-channel -- this term describes the
photon interaction with the fermion exchange current.

Let us return to Eq. (\ref{30}) and rewrite it in the form
similar to (\ref{11}).
One can see that
\be
\label{32}
i\;m\left(\hat H_0\hat r_\alpha-\hat r_\alpha\hat H_0\right)\ =\
\hat d_\alpha,
\ee
where $\hat H_0$ is the Hamiltonian
for composite systems written in the left-hand side
(\ref{23}), and the operator $\hat
r_\alpha$ is determined as
\be
\label{33}
\hat r_\alpha\ =\ \left(\begin{array}{ccc} r_\alpha& , & 0\\
0 &, & 0 \end{array}\right).
\ee
Substituting Eq. (\ref{32}) to (\ref{30}), we have
\be
\label{34}
F^{\phi\to\gamma f_0}_{\mu\alpha}\ =\
\int
d^3r\mbox{ Sp}_2\left[(\mbox{\boldmath$\sigma\cdot r$})
\psi_{f_0(s\bar s)}(r)
r_\alpha\sigma_\mu\psi_{\phi (s\bar s)}(r)\right]
\,  2i\;m(\varepsilon_\phi-
\varepsilon_{f_0}).
\ee
This formula for the dipole emission of photon is similar to that
of (\ref{11}) for the one-channel model.

\subsubsection{Partial width of the decay $\phi \to\gamma f_0$}

$\;$Partial width of the decay $\phi \to f_0$ in  case,
when $\phi $ is pure $s\bar s$ state, is determined by the following
formula:
\be
\label{pw1}
m_\phi \Gamma_{\phi \to \gamma f_0} =\frac{1}{6}\alpha
\frac{m^2_\phi - m^2_{f_0}}{m^2_\phi}
\left |A_{\phi \to  \gamma f_0(s\bar s)}\right |^2,
\ee
where $\alpha =1/137$ and the
$A_{\phi \to  \gamma f_0(s\bar s)}$
amplitude
is determined by Eq. (\ref{5a}) (here it is specified that we deal with
$s\bar s$ quarks in the intermediate state).

\subsection{Three-component model  ($s\bar s$, $n\bar n$, $gg$)
for $f_0$ and $\phi$}

$\;$ The above formula can be easily generalized for the
case, when $f_0$ is the three-component system ($s\bar s$, $n\bar n$,
$gg$) and $\phi$ is  two-component one ($s\bar s$, $n\bar n$), while
$gg$ is supposed to be
negligibly small. We have two transition form factors:
\be
\label{52a}
F^{\phi\to\gamma f_0(s\bar s)}_{\mu\alpha}\ =\
 \int
d^3r\mbox{ Sp}_2\left[(\mbox{\boldmath$\sigma r$})
\psi_{f_0(s\bar s)}(r)
r_\alpha\sigma_\mu\psi_{\phi (s\bar s)}(r)\right]
\, 2i\;m(\varepsilon_\phi-
\varepsilon_{f_0})
\ee
and
\be
\label{52b}
F^{\phi\to\gamma f_0(n\bar n)}_{\mu\alpha}\ =\
 \int
d^3r\mbox{ Sp}_2\left[(\mbox{\boldmath$\sigma r$})
\psi_{f_0(n\bar n)}(r)
r_\alpha\sigma_\mu\psi_{\phi (n\bar n)}(r)\right]
\, 2i\;m(\varepsilon_\phi-
\varepsilon_{f_0}).
\ee
The partial width reads
\be
\label{52c}
m_\phi \Gamma_{\phi \to \gamma f_0} =\frac{1}{6}\alpha
\frac{m^2_\phi - m^2_{f_0}}{m^2_\phi}
\left |A_{\phi\to\gamma f_0(s\bar s)}
+A_{\phi\to\gamma
f_0(n\bar n)}\right |^2 \, ,
\ee
with  $A_{\phi\to\gamma f_0}$ defined by Eqs. (\ref{4}) and
(\ref{5a}).
The charge factors, which were separated in
Eq. (\ref{4}), are equal to:
\be
Z^{(s\bar s)}_{\phi\to\gamma f_0}
=-\frac 23\, , \qquad Z^{(n\bar n)}_{\phi\to\gamma f_0} =\frac13\, ,
\ee
they include the combinatorics factor $2$ related to two diagrams
with the photon emmision by quark and antiquark,
see \cite{epja,s} for more detail.

\section{Decay $\phi(1020)\to\gamma f_0(980)$ }

$\;$ The vector meson $\phi(1020)$ has rather small decay width,
$\Gamma_{\phi(1020)} \simeq4.5\,$MeV; from this point of view there is
no doubt that treating $\phi(1020)$ as  stable particle
is  reasonable. As to $f_0(980)$, the picture is not so determinate. In
the PDG compilation
\cite{PDG}, the $f_0(980)$ width is given in the interval
$40\le\Gamma_{f_0(980)}\le100\,$MeV, and the width uncertainty
is related not to the data unaccuracy (experimental data are
rather good) but a vague definition of the width.

The mass and width of the resonance are determined by the pole position
in the complex-mass plane, $M=m-i\,\Gamma/2$, --- just this magnitude
 is a universal characteristics of the resonance.

\subsection{The $f_0(980)$: position of poles}

$\;$ The definition
of the $f_0(980)$ width is aggravated by the $K\bar K$
threshold singularity that leads to the existence of two, not one,
poles. According to the $K$-matrix analyses \cite{kmat,ufn}, there are
two poles in the  $(IJ^{PC}=00^{++})$ wave at $s\sim 1.0$ GeV$^2$,
\be
M^{\rm I} \simeq
1020-i40\mbox{ GeV }, \quad
M^{\rm II} \simeq  960-i200\mbox{ GeV },
\label{35}
\ee
which are
located on the different complex-$M$ sheets related to the
$K\bar K$-threshold, see Fig. 4. By switching off the decay
$f_0(980)\to K\bar K$, both poles begin to move to one another, and
they coincide after switching off the  $K\bar K$ channel completely.
Usually, when one discusses the $f_0(980)$, the resonance is
characterised by the closest pole, $M^{\rm I}$. However, when we are
interested in how far from each other the $\phi(1020)$ and $f_0(980)$
are, one should not forget about the second pole.

Keeping in mind the existence of two poles, one should
accept that $\phi(1020)$ and
 $f_0(980)$ are considerably "separated" from each other, and the
$f_0(980)$ resonance can hardly be represented as  stable particle --
we return to this point once more in Section 6 discussing the $\pi\pi$
spectrum in $\phi (1020)\to \gamma \pi\pi$.

\subsection{Switching off decay channels: bare states in
$K$-matrix analysis of the $(IJ^{PC}=00^{++}$) wave}

$\;$ A significant trait of the $K$-matrix analysis is that it also gives
us, along with the characteristics of real resonances, the positions of
levels before the onset of the decay channels, i.e. it determines the
bare states. In addition, the $K$-matrix analysis allows one to observe
the transform of bare states into real resonances. In Fig. 5,
one can see
such a transform of the $00^{++}$-amplitude poles
by switching off the decays
$f_0\to\pi\pi,K\bar K,\eta\eta,\eta\eta',\pi\pi\pi\pi$.
It is seen that, after switching off the decay channels, the
$f_0(980)$ turns into stable state, approximately  300 MeV lower:
\be
f_0(980)\ \longrightarrow\ f^{\rm bare}_0(700\pm100).
\label{36}
\ee

The transform of bare states into real resonances can be illustrated
by Fig. 6 for the levels in the potential well: bare states are the
levels
in a well with impenetrable wall (Fig. 6$a$); at the onset of the decay
channels (under-barrier transitions, Fig. 6$b$) the stable levels
transform into real resonance.

Figure 7 demonstrates
the evolution of coupling constants at the onset of the decay channels:
following \cite{content}, relative changes of the coupling constants
are shown for $f_0(980)$
after switching on/off the decay channels.
The onset of the decay channels is regulated
by the parameter $x$, and the value
$x=0$ corresponds to the bare state
(amplitude pole on the (Re $M$)-axis) and the value
$x=1$ stands for the resonance observed experimentally.

Let us bring the attention to a rapid increase of the coupling
constant $f_0\to K\bar K$ on the
evolution curve $f_0^{\rm bare}(700)$--$f_0(980)$
in the region $x\sim $ 0.8--1.0, where
$\gamma^2(x=1.0)-\gamma^2(x=0.8)\simeq 0.2$, see Fig. 7.
Actually this increase allows one to estimate a
possible admixure of the long-range $K\bar K$
component in the $f_0(980)$: it cannot be greater than $20\%$.

\subsection{ Calculation of the decay amplitude $\phi(1020)\to\gamma
f_0(980)$}

$\;$ The above discussion of the location of the amplitude poles of
$f_0(980)$ as well as the movement of poles by switching off the decay
channels tell us definitely that the smallness of the amplitude of the
$\phi(1020)\to\gamma f_0(980)$ decay due to a visible proximity of
masses of vector and scalar particles is rather questionable. As to
 $f_0(980)$, its poles "dived" into complex plane, in the average
in $\sim$100 MeV
(40~MeV for one pole and 200~MeV for another). But when we
intend to represent $f_0(980)$ as a stable level,
one should bear in mind that the mass of
the stable level is below the mass of  $\phi(1020)$ in $\sim300\,$MeV
--- this value is given by the $K$-matrix analysis. In both cases we
deal with the shifts in mass scale of the order of pion mass, that is
hardly small in hadronic scale.

The $K$-matrix amplitude of the  $00^{++}$ wave reconstructed in
\cite{kmat} gives us the possibility to trace the evolution of the
transition form factor
$\phi(1020)\to\gamma f^{\rm bare}_0(700\pm100)$ during the transformation of
 the bare state  $f^{\rm bare}_0(700\pm100)$ into the $f_0(980)$ resonance.
 Using the diagrammatic language, one can say that the
evolution of the form factor
$F^{(bare)}_{\phi\to\gamma f_0}$ is due to the processes shown in Fig.
8: $\phi$ meson goes into $f^{\rm bare}_0(n)$, with the emission of a
photon, then $f^{\rm bare}_0(n)$ decays into mesons
$f^{\rm bare}_0(n)\to hh=\pi\pi$,
$K\bar K$, $\eta\eta$, $\eta\eta'$, $\pi\pi\pi\pi$. The decay
yields may rescatter thus coming to  final states

The residue of the amplitude pole
$\phi(1020)\to\gamma\pi\pi$ gives us the transition amplitude
$\phi(1020)\to\gamma f_0(980)$. So,
in the $K$-matrix representation the amplitude of the reaction
$\phi(1020)\to\gamma\pi\pi$, Fig. 8, reads
\be
\label{37}
A_{\phi(1020)\to\gamma\pi\pi}(s)=\sum_{a,n}
\frac{F^{(bare)}_{\phi (1020)
\to\gamma f_0^{\rm bare}(n)}}{M^2_n-s}\,g^{\rm bare}_a(n)\left(
\frac1{1-i\hat\rho(s)\hat K(s)}\right)_{a,\pi\pi}.
\ee
Here $M_n$ is the mass of bare state, $g^{\rm bare}_a(n)$ is the coupling
for the transition
$f^{\rm bare}_0(n)\to a$, where $ a=\pi\pi$, $K\bar K$, $\eta\eta$,
$\eta\eta'$, $\pi\pi\pi\pi$. The matrix element
$(1-i\hat\rho(s)\hat K(s))^{-1}$ takes account of the rescatterings
of the formed mesons. Here $\hat\rho(s)$ is the diagonal matrix
of phase spaces for hadronic states (for example, for the
$\pi\pi$ system it reads $\rho_{\pi\pi}(s)=\sqrt{(s-4m^2_\pi)/s}$),
and the  $K$-matrix elements
$K_{ab}(s)$ contain the poles corresponding to the bare states:
\be
K_{ab}(s)=\sum_n \frac{g^{\rm bare}_a(n) g^{\rm bare}_b(n)}{M^2_n-s}+f_{ab}(s).
\ee
The function
$f_{ab}(s)$ is analytical in the right-hand half-plane of the
complex-$s$ plane,
at Re$s>0$, see \cite{kmat} for more detail.

Near the pole corresponding to $f_0$ resonance
(resonance poles are contained in the factor
$(1-i\hat\rho(s)\hat K(s))^{-1}$ ),
the amplitude $\phi(1020)\to\gamma\pi\pi$ is written as follows:
\be
\label{38}
A_{\phi(1020)\to\gamma\pi\pi}(s)\ \simeq\
\frac{A_{\phi(1020)\to\gamma f_0(980)}}{M^2_{f_0(980)}-s}\,
g_{f_0(980)\to\pi\pi} + \mbox{ smooth
terms },
\ee
where $M_{f_0}(980)$ is the complex-valued resonance mass:
$M_{f_0(980)}\to M^{\rm I} \simeq 1020-i40$
MeV for the first pole, and $M_{f_0(980)}\to M^{\rm II}\simeq
960-i200$
MeV for the second one. The transition amplitude
$A_{\phi(1020)\to\gamma f_0(980)}$ is different for different poles;
the $g_{f_0(980)\to\pi\pi}$ couplings
are different as well.

We see that the radiative transition  $\phi (1020)\to
\gamma f_0(980)$ is determined by two amplitudes,\\
$A_{\phi(1020)\to\gamma f_0(M^{\rm I})}
\equiv A^{\rm I}_{\phi(1020)\to\gamma f_0(980)}$ and
$A_{\phi(1020)\to\gamma f_0(M^{\rm II})}\equiv
A^{\rm II}_{\phi(1020)\to\gamma f_0(980)}$,  and
just these amplitudes are the subject of our
interest. The amplitudes
$A^{\rm I}_{\phi(1020)\to\gamma f_0(980)}$,
$A^{\rm II}_{\phi(1020)\to\gamma f_0(980)}$ may be represented as the
sum of contributions from different bare
states:
\be
\label{39}
A^{\rm I}_{\phi(1020)\to\gamma
f_0(980)}=\sum_n \zeta^{\rm I}_n[f_0(980)]
F^{(bare)}_{\phi(1020)\to\gamma
f^{\rm bare}_0(n)},
\ee
$$
A^{\rm II}_{\phi(1020)\to\gamma
f_0(980)}=\sum_n
\zeta^{\rm II}_n[f_0(980)]F^{(bare)}_{\phi(1020)\to\gamma f^{\rm bare}_0(n)},
$$
To calculate the constants $\zeta_n[f_0(m_R)]$ we use the
$K$-matrix solution for the  $00^{++}$ wave amplitude denoted in
\cite{kmat} as II-2. In this solution, there are five bare states
$f_0^{\rm bare}(n)$
in the mass interval 290--1950~MeV: four of them are members of the
$q\bar q$ nonets ($1^3P_0q\bar q$ and $2^3P_0q\bar q$)
and the fifth state is the glueball. Namely:
\begin{eqnarray}
1^3P_0q\bar q: && f^{\rm bare}_0(700\pm 100),\quad
f^{\rm bare}_0(1220\pm30), \\ 2^3P_0q\bar q: &&
f^{\rm bare}_0(1230\pm40), \quad f^{\rm bare}_0(1800\pm40),
\nonumber\\ {\rm
glueball}: && f^{\rm bare}_0(1580\pm50). \nonumber \label{40}
\end{eqnarray}
For the first pole of the $f_0(980)$ resonance located at
$M[f_0(980)]=1020-i40\,$MeV the renormalization
constants are as follows:
\begin{eqnarray}
\zeta^{\rm (I)}_{700}[f_0(980)] &=& 0.62\exp(-i144^\circ),
\\
\zeta^{\rm (I)}_{1220}[f_0(980)] &=& 0.37\exp(-i41^\circ),
\nonumber\\
\zeta^{\rm (I)}_{1230}[f_0(980)] &=& 0.19\exp(i1^\circ), \nonumber\\
\zeta^{\rm (I)}_{1800}[f_0(980)] &=& 0.02\exp(-i112^\circ), \nonumber\\
\zeta^{\rm (I)}_{1580}[f_0(980)] &=& 0.02\exp(i5^\circ).\nonumber
\label{41}
\end{eqnarray}
These constants are complex-valued. One should pay attention to the fact
that the phases of constants $\zeta^{\rm (I)}_{700}[f_0(980)]$ and
$\zeta^{\rm (I)}_{1220}[f_0(980)]$ have the relative  shift close to
$90^\circ$. This means that the contributions
from $f^{\rm bare}_0(700\pm 100)$
and $f^{\rm bare}_0(1220\pm 30)$ (which are members of the basic
$1^3P_0q\bar q$ nonet) do not interfere practically in the calculation
of probability for the decay $\phi(1020)\to\gamma f_0(980)$.

Actually
one may neglect the bare states
$f^{\rm bare}_0(1230)$, $f^{\rm bare}_0(1800)$, $f^{\rm bare}_0(1580)$ in the
calculation of the $\phi(1020)\to\gamma f_0(980)$ reaction, bacause
the form factors for the production of radial excited states are
noticeably suppressed, see \cite{s}:
$$
\left|F^{(bare)}_{\phi(1020)\to\gamma f_0(2^3P_0q\bar q)}\right| \ll
\left|F^{(bare)}_{\phi(1020)\to\gamma f_0(1^3P_0q\bar q)}\right| . $$
Besides, the coefficients
$\zeta^{\rm (I)}_{1230}[f_0(980)]$
$\zeta^{\rm (I)}_{1800}[f_0(980)]$ are also
comparatively small, see (\ref{41}).

The second pole located on the third sheet,
$M[f_0(980)]=960-i200\,$MeV, has  renormalizing constants as follows:
\begin{eqnarray}
\zeta^{\rm (II)}_{700}[f_0(980)] &=& 1.00\exp(i6^\circ), \nonumber\\
\zeta^{\rm (II)}_{1220}[f_0(980)] &=& 0.33\exp(i113^\circ), \nonumber\\
\zeta^{\rm (II)}_{1230}[f_0(980)] &=& 0.32\exp(i148^\circ), \nonumber\\
\zeta^{\rm (II)}_{1800}[f_0(980)] &=& 0.08\exp(i4^\circ), \nonumber\\
\zeta^{\rm (II)}_{1580}[f_0(980)] &=& 0.04\exp(i98^\circ).
\label{42}
\end{eqnarray}
Here, as before, the transitions
$\phi(1020)\to\gamma f^{\rm bare}_0(1230),\, \gamma f^{\rm bare}_0(1580),\,
\gamma f^{\rm bare}_0(1800)$  are negligibly small.

In $\phi(1020)$, the admixture of the $n\bar n$-component is small. In
the estimates given below we assume $\phi(1020)$ to be pure $s\bar
s$ state. The bare states $f_0^{\rm bare}(700)$ and $f^{\rm bare}_0(1220)$
are mixtures of the $n\bar n$ and $s\bar s$ components
$$
n\bar n\cos\varphi+s\bar s\sin\varphi, $$
and, according to \cite{kmat},
the mixing angles are as follows:
\begin{eqnarray}
\varphi\left[f^{\rm bare}_0(700)\right] &=& -70^\circ\pm10^\circ,
\nonumber\\
\varphi\left[f^{\rm bare}_0(1220)\right] &=& 20^\circ\pm10^\circ.
\label{43}
\end{eqnarray}
Because of that the transition amplitude
for $\phi(1020)\to\gamma f_0(980)$ reads
\begin{eqnarray}
A^N_{\phi(1020)\to\gamma f_0(980)} &\simeq& \zeta^N_{700}
[f_0(980)]\sin\varphi
\left[f^{\rm bare}_0(700)\right]
F^{(bare)}_{\phi(1020)\to\gamma f^{\rm bare}_0(700)}
\nonumber\\
&+&
\zeta^N_{1220}[f_0(980)]\sin\varphi\left[f^{\rm bare}_0(1220)\right]
F^{(bare)}_{\phi(1020)\to\gamma f^{\rm bare}_0(1220)}\, .
\label{44}
\end{eqnarray}
Here $\zeta^N_{700}[f_0(980)]$ and $\zeta^N_{1220}[f_0(980)]$
($N=I,II$) are given by the formulae
(\ref{41}), (\ref{42}). One can see that numerically the factor
$\zeta_{1220}[f_0(980)]\sin\varphi\left[f^{\rm bare}_0(1220)\right]$
is small, and we
may neglect the second term   in the right-hand
side  (\ref{44}). Then for the pole, which is the closest
one to the
real axis (1020 - i40 MeV), one  has:
\be
A^{\rm I}_{\phi(1020)\to\gamma f_0(980)}
\simeq(0.58\pm0.04)F^{(bare)}_{\phi(1020)\to\gamma
f_0^{\rm bare}(700)}\, ,
 \label{45}
\ee
and for the distant one, (960 - i200 MeV):
\be
\label{45a}
A^{\rm II}_{\phi(1020)\to\gamma
f_0(980)}
\simeq(0.92\pm0.06)F^{(bare)}_{\phi(1020)\to\gamma f_0^{\rm bare}(700)}\, .
\ee
We see that practically the $A^{\rm II}_{\phi(1020)\to\gamma f_0(980)}$
amplitude
 does not change
its value during the evolution from  bare state to resonance, while the
decrease of  $A^{\rm I}_{\phi(1020)\to\gamma f_0(980)}$ is significant.

\subsection{Comparison to data}

$\;$ Comparing  the above-written formulae to experimental data
we have parametrized the wave functions of the $q\bar q$ states in the
simplest, exponent-type, form (see Section 2.3). For
$\phi(1020)$, we accept  its mean radius square to be close to the
pion radius, $R^2_{\phi(1020)} \simeq R^2_{\pi}$:
both states are members of the same 36-plet. This value of the mean
radius square  for $\phi(1020)$ fixes its wave function by
$b_\phi= 10$ GeV$^{-2}$.

For $f_0^{\rm bare}(700)$, we change the value $b_{f_0}$ in the
interval
$$
5\, {\rm GeV}^{-2}\le b_{f_0}^{(bare)}\le 15\, {\rm
GeV}^{-2}\,
$$
that corresponds to the interval (0.5--1.5)$R^2_\pi$
of the mean radius square of $f_0^{(bare)}(700)$.

Using the branching ratios \cite{nov1,nov2} as follows:
\be
{\rm BR}[\phi(1020)\to\gamma
f_0(980)]= (3.5\pm 0.3^{\, +\, 1.3}_{\, -\, 0.5} )\times 10^{-4}\, ,
\label{54-1}
\ee
$$
{\rm BR}[\phi(1020)\to\gamma f_0(980)]=
(2.90\pm 0.21^{\pm 1.54})\times 10^{-4}\, ,
$$
and the definition of the radiative decay width:
$$
m_\phi \Gamma_{\phi \to \gamma f_0} =\frac{1}{6}\alpha
\frac{m^2_\phi - m^2_{f_0}}{m^2_\phi}
\left |A_{\phi \to
\gamma f_0}\right |^2 \, ,
$$
we have the following experimental value for the decay
amplitude:
\be
\label{54-2}
A^{\rm (exp)}_{\phi(1020)\to\gamma f_0(980)} =0.115\pm 0.040\,
{\rm GeV}\, .
\ee
Here  $\alpha =1/137$, $m_\phi =1.02$ GeV and
$m_{f_0} =0.975$ GeV (the mass reported in \cite{nov1,nov2}
for the measureed
$\gamma f_0(980)$ signal) and
$\Gamma_{\rm tot}[\phi(1020)]=4.26\pm 0.05$ MeV \cite{PDG}.
The right-hand side (\ref{54-2})
should be compared with $A_{\phi(1020)\to\gamma f_0(980)}$
calculated with Eqs. (\ref{11d}), (\ref{34}), and (\ref{45}):
\be
\label{54-3}
A_{\phi(1020)\to\gamma f_0(980)}^{\rm (calc)}({\rm dipole}) \simeq
(0.58\pm0.04) \sqrt{W_{q\bar q}[f^{\rm bare}_0(700)]}\,
Z^{(s\bar s)}_{\phi\to\gamma f_0}\times
\ee
$$
\times
\frac{2^{7/2}}{\sqrt3}\,
\frac{b^{7/4}_\phi b^{5/4}_{f_0}}{(b_\phi+b_{f_0})^{5/2}}\,
 m_s\left [ m_\phi-(0.7\pm 0.1){\rm GeV}\right ].
$$
Recall, in (\ref{54-3}) the factor $(0.58\pm0.04)$ takes into account
the change of the transition amplitude caused by the final-state
hadron interaction, Eq. (\ref{45}). The probability to find
quark--antiquark component
in the bare state $f^{\rm bare}_0(700)$ is denoted as
$W_{q\bar q}[f^{\rm bare}_0(700)]$: one can guess that it is of the order of
80--90\%, or even more. The mass of the strange constituent quark
is equal to
$m_s \simeq 0.5$ GeV. The wave functions of $\phi (1020)$
and $f^{\rm bare}_0(700)$ are parametrized as exponents:
we fix $b_\phi =10$ GeV$^{-2}$ (that gives for the mean radius
 of $\phi (1020)$ the value of the order of the pion radius
, $R_\phi  \simeq R_\pi$), and vary $b_{f_0}$ in the
interval (5--15) GeV$^{-2}$.

The comparison of the data (\ref{54-2}) to the calculated
amplitude is shown in Fig. 9. We see that the calculated
amplitude (\ref{54-3}) is in a perfect agreement with data, when
$M_{f_0^{(bare)}}\simeq $750--800 MeV, that is just inside the error
bars given by the $K$-matrix analysis \cite{kmat}.

\section{Pion-pion spectrum in $\phi (1020)\to \gamma \pi\pi$}

$\;$ The $f_0(980)$ resonance is seen in the reaction
$\phi(1020)\to \gamma\pi\pi$
as a peak at the edge of the $\pi\pi$ spectrum. So it
is rather enlightening to calculate the $\pi\pi$
spectrum to be sure that its description agrees both with the quark
model calculation of the form factor $F_{\phi(1020)\to \gamma
f_0(980)}$
and  threshold theorem (cross section tending to zero as
$\omega^3$ at $\omega\to 0$, where $\omega=m_\phi-M_{\pi\pi}$).

Partial cross section of the decay $\phi(1020)\to \gamma\pi^0\pi^0$ is
given by the following formula:
\be
\label{sp1}
\frac{d\Gamma_{\phi (1020)\to\gamma \pi^0\pi^0} }{dM_{\pi\pi }}=
\frac13\Gamma_{\phi (1020)\to\gamma f_0(980)}\,
\frac{m_\phi ^2-M^2_{\pi\pi}}{m_\phi ^2-m^2_{f_0}}\,
\times
\ee
$$
\times
\frac {2M_{\pi\pi}}{\pi}\rho_{\pi\pi}
\left |\frac{g_\pi}{M^2_0-M^2_{\pi\pi} -ig_\pi^2 \rho_{\pi\pi}
-ig_K^2\rho_{K\bar K}} +B(M^2_{\pi\pi} ) \right |^2 \, .
$$
The factor $1/3$ in front of the right-hand side  (\ref{sp1})
is associated with the $\pi^0\pi^0$ channel:
$\Gamma_{\phi (1020)\to\gamma \pi^0\pi^0}= 1/3
\Gamma_{\phi (1020)\to\gamma \pi\pi} $. Here for the description of the
$f_0(980)$ we use the Flatt\'e formula
\cite{Flatte} with the phase space factors
\be
\label{sp2}
\rho_{\pi\pi}=\frac1M_0
\sqrt{M^2_{\pi\pi}-4m^2_\pi}\, ,\quad
\rho_{K\bar K}=\frac1M_0\sqrt{M^2_{\pi\pi}-4m^2_K}.
\ee
At $M^2_{\pi\pi}<4m^2_K$ one should replace
$\sqrt{M^2_{\pi\pi}-4m^2_K}\to i \sqrt{4m^2_K-M^2_{\pi\pi}}$.
In line with \cite{nov1,nov2,sarantsev}, we use  the
Flatt\'e formula with the parameters
\be
\label{sp4}
g_\pi^2=0.12 \, {\rm GeV}^2 \, ,\quad
g_K^2=0.27 \, {\rm GeV}^2 \, , \quad M_0=0.975\, {\rm GeV} \, .
\ee
The threshold theorem requires
\be
\label{sp5}
\left [\frac{g_\pi}{M^2_0-M^2_{\pi\pi} -ig_\pi^2
\rho_{\pi\pi}
-ig_K^2\rho_{K\bar K}} +B(M^2_{\pi\pi})  \right ]_
{M_{\pi\pi} \to m_\phi} \sim (M_{\pi\pi} - m_\phi) ,
\ee
that gives a constraint for the background term $B(M^2_{\pi\pi})$.
The term
$B(M^2_{\pi\pi})$ is parametrized in the form:
\be
\label{sp6}
B(M^2_{\pi\pi}  )
= C\left [1+a(M^2_{\pi\pi} - m^2_\phi)\right ]
\exp{\left [-\frac{ m^2_\phi -M^2_{\pi\pi}}{\mu^2}\right ]},
\ee
and the  parameter $C$ is fixed by the constraint:
\be
\label{sp7}
\left [\frac{g_\pi}{M^2_0-M^2_{\pi\pi} -ig_\pi^2
\rho_{\pi\pi}
-ig_K^2\rho_{K\bar K}} +C  \right ]_
{M_{\pi\pi} = m_\phi} =0\, .
\ee
Fitting to the $\pi^0\pi^0$ spectrum \cite{nov1}, see Fig. 10$a$,
we have the following values for other parameters:
\be
\label{sp7a}
\frac1a=-0.2\, {\rm GeV}^{2}, \quad \mu=0.388\, {\rm GeV}\, .
\ee
For $\Gamma_{\phi(1020)\to \gamma f_0(980)}$ entering (\ref{sp1})
we have used $A_{\phi(1020)\to \gamma f_0(980)}=0.13$ GeV that satisfies
both (\ref{54-2}) and (\ref{54-3}).

The Flatt\'e formula gives us rather rough
description of the $\pi\pi$ amplitude around
the $f_0(980)$ resonance. More precise description may be obtained
by using  in addition
the non-zero transition length for $\pi\pi\to K\bar K$
\cite{content}.  For this case, we have the formulae analogous to
Eq. (\ref{sp1}), after replacing the resonance factor
\be
\label{sp8}
\frac{g_\pi}{M^2_0-M^2_{\pi\pi} -ig_\pi^2 \rho_{\pi\pi}
-ig_K^2\rho_{K\bar K}}
\ee
by the following one:
\be
\label{sp9}
\frac{g_\pi+i\rho_{K\bar K} g_Kf}{M^2_0-M^2_{\pi\pi} -ig_\pi^2
\rho_{\pi\pi} -i\rho_{K\bar K}[g_K^2+i\rho_{\pi\pi}(2g_\pi g_Kf
+f^2(M_0^2-M^2_{\pi\pi}))]}\, .
\ee

The parameters  found in \cite{content}
are equal to:
\begin{eqnarray}
g_\pi=0.386\ {\rm GeV}, && g_K=0.447\ {\rm GeV}, \\
M_0=0.975\ {\rm GeV} , && f=0.516.\nonumber
\label{sp10}
\end{eqnarray}
The transition length $a_{\pi\pi\to K\bar K}$ is determined
by the parameter $f$ as follows:
$a_{\pi\pi\to K\bar K}=2f/M_0$.

The description of the $\pi^0\pi^0$ spectra
\cite{nov1} within the resonance
formulae (\ref{sp9}) is demonstrated in Fig. 10$b$. In this fit we have
the following parameters for $B(M^2_{\pi\pi})$:
\be a=0\, , \quad \mu=0.507\, {\rm GeV}.
\ee
In this variant of the fitting to spectra we also used
$A_{\phi(1020)\to \gamma f_0(980)}=0.13$ GeV .

Let us emphasize that the visible width of the $f_0(980)$
signal in the $\pi\pi$ spectrum is comparatively large, $\sim$150 MeV,
that is related  to an essential contribution of the
second pole at $960-i200$ MeV.

\section{The additive quark model, does it work?}

$\;$ Let us point to the two aspects of this question. One is the
problem of the applicability of the additive quark model to the
production of the resonance $f_0(980)$, another one is  the
production of the bare state $f_0^{\rm bare}(700\pm 100)$.

\subsection{ Process $\phi(1020)\to\gamma f_0^{\rm bare}(700\pm 100)$}

The additive quark model describes well the production  of the bare
state $f_0^{\rm bare}(700\pm 100)$, provided its mass is in the region
750--800 MeV. To see it, consider Eq. (\ref{34}) for
$F^{\phi\to\gamma f_0}_{\mu\alpha}$(dipole),
or Eq. (\ref{11d}), where exponential
representation of the quark wave functions is used. Formula  (\ref{11d})
 takes into account
both the additive quark model processes and photon emission by the
charge-exchange current, while Eq. (\ref{11c}) gives us the
triangle-diagram contribution within  additive quark model. The
contribution of the charge-exchange current is small, when
\be
m_s[m_\phi-M_{f_0^{\rm bare}}]\simeq \frac1{b_\phi}.
\label{A.1}
\ee
At $m_s=0.5$ GeV and $b_\phi=10$ GeV$^{-2}$ the equality
(\ref{A.1}) is
almost fulfilled, when $M_{f_0^{\rm bare}}\simeq 0.8$ GeV.
Such a magnitude is allowed by the
$K$-matrix fit \cite{kmat}, which gives $M_{f_0^{\rm bare}}=0.7\pm0.1$
GeV.

However, let us emphasize that
the error bars $\pm 0.1$ GeV are rather large in the difference
$(m_\phi-M_{f_0^{\rm bare}})$: with the lower possible limit
$M_{f_0^{\rm bare}}=0.6$ GeV we face a two-times disagreement in Eq.
(\ref{A.1}).
Still, one may hardly hope that the $K$-matrix analysis of the
$00^{++}$ wave would provide us with a tighter restriction for the
mass of this  bare state, since
a large
uncertainty in the definition of $M_{f_0^{\rm bare}}$ is not
related to the data accuracy but to the problem of  the
light $\sigma$ meson existence, see the discussions in
%\cite{ufn,tuan,vanbeveren,minkowski}
[5,26--28] and references therein.

The use of $F^{\phi\to \gamma f_0}_{\mu\alpha}$(additive),
Eq. (\ref{11c}),
 for the calculation of
$A^{(calc)}_{\phi(1020)\to \gamma f_0(980)}$ results in the
agreement with experimental data. Thus we have:
\be
A_{\phi(1020)\to\gamma f_0(980)}^{(calc)}{\rm (additive)}\simeq
(0.58\pm0.04) \sqrt{W_{q\bar q}[f^{\rm bare}_0(700)]}\,
Z^{(s\bar s)}_{\phi\to\gamma f_0}\times
\ee
$$
\times
\frac{2^{7/2}}{\sqrt3}\,
\frac{b^{3/4}_\phi b^{5/4}_{f_0}}{(b_\phi+b_{f_0})^{5/2}}\, .
$$
In Fig. 11, one can see
$A^{(calc)}_{\phi(1020)\to \gamma f_0(980)}{\rm (additive)}$
 versus
$A^{\rm (exp)}_{\phi(1020)\to \gamma f_0(980)}$: there is a good
agreement with data.

The existence of two characteristic sizes in a hadron, namely,
hadronic
radius and that of constituent quark, may be the reason why the
contribution of the charge-exchange current is  small in the
reaction $\phi(1020)\to \gamma f_0^{\rm bare}(700)$. Relatively small
radius of the constituent quark assumes that charge-exchange
interaction $s\bar s\to gg\to n\bar n$ is a short-range one, that
causes a smallness of the second term in the right-hand side
(\ref{30a}).

The hadronic size is defined by the confinement radius
$R_h\sim R_{\rm conf}$, which is of the order of 1 fm for light
hadrons. The constituent quark size, $r_q$, is much smaller, it is
defined, as one may believe, by relatively large mass of the soft gluon
(experimental data \cite{glue1} and lattice calculations \cite{glue2}
give us $m_g\sim $700--1000 MeV). So we get
$r^2_q/R^2_h\sim $(0.1--0.2), the same value follows from the analysis
of soft hadron collisions, see \cite{book,dakhno} and references
therein.

\subsection{Process $\phi(1020)\to\gamma f_0(980)$}

The two sizes, $r^2_q$ and $R^2_h$, being accepted, the
additive quark model contribution
 dominates the
reaction  $\phi(1020)\to \gamma f_0(980)$ too, thus
allowing direct use of the triangle diagram of Fig. 1$a$
for the calculation of this process. Such calculations were performed in
\cite{s}, revealing  resonable agrement with data.
Once again it should be emphasized that the triange diagram
contribution does not have a particular smallness related to
a deceptive proximity of $\phi(1020)$ and $f_0(980)$. Besides, as was
explained above, the  poles associated with these
resonances are separated from each other in the complex-$M$ plane in
nonsmall distances in the hadronic scale.

In the literature there exist rather opposite statements about the
possibility to describe the reaction $\phi(1020)\to \gamma f_0(980)$
within the frame of the hypothesis of the $q\bar q$ nature of
$f_0(980)$. Using the QCD sum-rule technique the authors of
\cite{penn-f} evaluated the rate of the decay
$\phi(1020)\to \gamma f_0(980)$, with a fair agreement  with data,
supposing a sizeable $s\bar s$ component in the $f_0(980)$.

The results of the calculation performed in \cite{deWitt} in the
framework of the additive quark model do not agree with data on
the reaction $\phi(1020)\to \gamma f_0(980)$. This calculation though
similar to those of \cite{epja,s} led to different result, so it would
be instructive to compare  model parameters used in these two
approaches.

In \cite{deWitt} as well as \cite{epja,s}, the exponential
parametrization of the wave function was used, however the slopes
$b_{\phi}$ and $b_{f_0}$  in \cite{deWitt} were considerably smaller
(constituent quark masses are smaller too). In \cite{deWitt},
 $b_{s\bar s}= 2.9$ GeV$^{-2}$ and
$b_{u\bar u}=b_{d\bar d}= 3.7$ GeV$^{-2}$ ($m_u=m_d=220$ MeV, $m_s=450$
MeV), while in  \cite{epja,s}  $b_\phi\simeq 10$ GeV$^{-2}$ and
$b_\phi\sim b_{f_0}$ ($m_u=m_d=350$ MeV, $m_s=500$ MeV). Besides, in
\cite{deWitt} the scheme of the mixing of  $f_0$ states
 was used  that was suggested in \cite{LW,CK}, where the transitions
$f_0^{\rm bare}\to$ real mesons were not accounted for. Still, as was
emphasized above (Section 5.2), just the transitions
$f_0^{\rm bare}\to \pi\pi,K\bar K,\eta\eta,\pi\pi\pi\pi$ afford the final
disposition of poles in the complex plane, for they are responsible for
the resonance mass shift of the order of 100 MeV, see Fig. 5.

In our opinion, the failure of the $q\bar q$ model demonstrated  in
\cite{deWitt}  can testify only the fact that not any
model enables the description of  radiative decays. The $q\bar
q$ model should be based on the whole set of experimental data but not
on the reproducing several levels of the lowest states.

 \section{Conclusion}

$\;$ Correct determination of the origin of $f_0(980)$ is a key
for understanding of the status
of the light $\sigma$ and classification of heavier mesons
$f_0(1300)$, $f_0(1500)$, $f_0(1750)$ and the broad state
$f_0$(1200--1600).

We have shown that experimental data on the reaction
$\phi(1020)\to \gamma f_0(980)$ do not contradict the suggestion about
the dominance of the
quark--antiquark component in the $f_0(980)$. However, as was
emphasized in Introduction, the final conclusion about the origin of
$f_0(980)$ should be made on the basis of the whole availability
of arguments, so let us enumerate them briefly.

1. There are data on the hadronic decays of the lowest
mesons, and the most reliable information on scalar resonances is given
by the $K$-matrix analysis.  Summing up, one can state that, according
to the $K$-matrix analysis
\cite{kmat},  the lowest states $f_0(980)$, $f_0(1300)$,
$a_0(980)$, $K_0(1430)$ are the descendants of bare mesons, which have
created the $1^3P_0$ multiplet. Because of that, all the decays,
namely,
\begin{eqnarray}
f_0(980)&\to& \pi\pi, K\bar K, \\
f_0(1300)&\to &\pi\pi, K\bar K, \eta\eta, \nonumber \\
a_0(980)&\to &K\bar K,\pi\eta,              \nonumber \\
K_0(1430)&\to & K\pi   \nonumber
\end{eqnarray}
are described by two parameters only in the leading terms of the $1/N_c$
expansion, which are  the universal coupling constant and mixing angle
for the $n\bar n$ and $s\bar s$-components in the scalar--isoscalar
sector.

2. Another argument is  the systematics of scalar states on
linear trajectories on the $(n,M^2)$ plane. For scalar mesons the
trajectories are shown in Fig. 12$a$ --- one can see that $f_0(980)$
lays comfortably on linear trajectory, together with the other scalars.
Such trajectories are formed not only for the scalar sector but also for
all aggregate of data, see \cite{syst,ufn}, and all the
trajectories are characterized by the universal slope.

Similar $q\bar q$ trajectories exist in the $(J,M^2)$-plane too, and
$f_0(980)$ belongs to one of them.

3. The alternative to the $q\bar q$ system may be the four-quark
$q\bar q q\bar q$, or molecular $K\bar K$ structure
\cite{jaffe,isgur,close,achasov}. Such a nature   of
$f_0(980)$ would mean that $f_0(980)$ was loosely bound
system but the experiment tells us that this is not so. The
matter is that

(i) $f_0(980)$ is easily produced at large momenta transferred to the
nucleon in the reaction $\pi^- p\to f_0(980) n$ \cite{gams,bnl},

(ii)  $f_0(980)$ is produced in the $Z^0$-boson decays \cite{chlap},

(iii) $f_0(980)$ is produced in central $pp$ collisions at high
energies \cite{pp}.

Were the $f_0(980)$ a loosely bound
system, these processes would be suppressed.

4. The $f_0(980)$ resonance is produced in hadronic decays of the $D_s$
meson, $D^+_s\to \pi^+ f_0(980)$, with the probability comparable with
that for the transition $D_s^+\to \pi^+\phi(1020)$ \cite{Ds}.
These two reactions
are due to the weak decay of $c$ quark, $c\to \pi^+ s$; as to
$f_0(980)$, it is formed, like $\phi(1020)$, by the $s\bar s$ system
in the transition $s\bar s \to f_0(980)$. The calculation of this
process \cite{weDs} shows us that the $f_0(980)$ yield in the reaction
$D_s^+\to \pi^+f_0(980)$ can be reliably calculated under the
assumption that $f_0(980)$  is close to the $q\bar q$ flavor octet.

The study of the $D_s^+\to \pi^+f_0(980)$ decay by
%\cite{ochs,van,gatto}
[44-46]
also led to the conclusion about the $s\bar s$ nature of $f_0(980)$.

5. Concerning radiative decays with the formation of   $f_0(980)$,
we see that the transition $\phi(1020)\to \gamma f_0(980)$ can
be well described within the approach of additive quark model,
with the dominant $q\bar q$ component in the $f_0(980)$. Another
radiative decay, $f_0(980)\to \gamma\gamma$, partial width of which
was measured \cite{pennington}, can be also treated in terms of the
$q\bar q$ structure of $f_0(980)$ \cite{epja,tensor}. The values of
partial widths in both decays support the conclusion made in
\cite{kmat} that the flavor content of $f_0(980)$ is close the octet
one.

We thank
Ya.I. Azimov, L.G. Dakhno, S.S. Gershtein, Yu.S. Kalashnikova,
A.K. Likhoded, M.A. Matveev,
V.A. Markov, D.I.  Melikhov, A.V. Sarantsev and W.B. von Schlippe for
helpful and stimulating discussions of problems involved.

This work is supported by the RFBR Grant N 04-02-17091.

\section*{Appendix A: Dipole emission of photon }

$\;$ To describe the interaction of composite system with
electromagnectic field, we consider
the full Hamiltonian:
\be
\label{a1}
\hat H(0)=\left|\begin{array}{ccc}
\frac{ k^2_1}{2m}+\frac{k^2_2}{2m} +
U_{s\bar s\to s\bar s}({\bf r}_1-{\bf r}_2) &, & \widehat
U_{s\bar s\to gg}({\bf r}_1-{\bf r}_2)\\
\widehat U_{s \bar s\to gg}({\bf r}_1-{\bf r}_2)& , &
 \frac{k^2_1}{2m_g}+\frac{k^2_2}{2m_g} +U_{gg\to gg}({\bf r}_1-{\bf
 r}_2)
\end{array}\right| .
\ee
Here the coordinates (${\bf r}_a$) and momenta
(${\bf k}_a=-i{\bf \nabla}_a$) of the constituents are related to the
characteristics of the relative movement, entering
 (\ref{23}), as follows
\be
{\bf r}_1 =\frac 12 {\bf r} +{\bf R} \, , \quad
{\bf r}_2 =-\frac 12 {\bf r} +{\bf R} \, ,     \quad
{\bf k}_1 =\frac 12 {\bf k} +{\bf P} \, , \quad
{\bf k}_2 =-\frac 12 {\bf k} +{\bf P} \, .
\ee
The electromagnetic interaction is included  by
substituting in (\ref{a1}) as follows:
\be
\label{a2}
{\bf k}^2_1 \to
\left ({\bf k}_1-e_1{\bf A}(r_1)\right )^2,\qquad
 {\bf k}^2_2 \to
\left ({\bf k}_2-e_2{\bf A}(r_2)\right )^2,
\ee
 $$
\widehat U_{s\bar s\to gg}({\bf r}_1-{\bf r}_2) \to
\widehat U_{s\bar s\to gg}({\bf r}_1-{\bf r}_2)
\exp{\left[ie_1\int\limits^{{\bf r}_1}_{-\infty}
dr'_\alpha A_\alpha (r')
+ie_{2}\int
\limits^{{\bf r}_2}_{-\infty}
dr'_\alpha A_\alpha (r')\right ]},
$$
with $e_1=-e_2=e_s$.
After that we obtain the gauge-invariant Hamiltonian $\hat H(A)$:
\be
\label{28}
\hat H({\bf A})\ =\ \hat\chi^+
\hat H({\bf A}+ {\bf\nabla}\chi)\hat\chi,
\ee
where ${\bf A}+ {\bf\nabla}\chi $ means the following
substitution:
\be
{\bf A}(r_a)\to{\bf A}(r_a)+{\bf\nabla}\chi(r_a)\, ,
\ee
and matrix $\hat\chi $ reads
\be
 \hat\chi\
=\left|\begin{array}{ccc} \exp[ie_s\chi(r_1)-ie_{ s}\chi(r_2)]& , &
0\\ 0\ & , & 1\end{array} \right|.
\ee
For the transition
$\phi\to\gamma f_0$, keeping the terms proportinal to the
$s$-quark
charge, $e_s$,
we have the following operator for the dipole emission:
\be \label{29a}
\hat d_\alpha\ =\ \left|\begin{array}{ccc}
2(k_{1\alpha}-k_{2\alpha})&, & i(r_{1\alpha}-r_{1\alpha})\hat U_{s\bar
s\to gg}({\bf r}_1-{\bf r}_2) \\
-i(r_{1\alpha}-r_{1\alpha}) \hat U_{s\bar s\to gg}({\bf r}_1-{\bf
r}_2)& , & 0 \end{array} \right|.
\ee

There exist other mechanisms of the photon emission which,
being beyond the additive quark model,
lead us to the dipole formula for $V \to \gamma S$ transition, an
example is given by
$({\bf L\cdot S})$-interaction in the quark--antiquark component \cite{8,9,yu}.
The short-range  ($\bf L\cdot S$) interaction in the $q\bar q$ systems,
was discussed in \cite{sz,glashow} as a source of the nonet splitting.
Actually the point-like $({\bf L\cdot S})$ interaction gives
$(v/c)$ corrections to the nonrelativistic approach. In
the relativistic quark model approaches based on the Bethe-Salpeter
equation, the gluon-exchange forces result in similar nonet splitting as
for the $({\bf L\cdot S})$ interaction, for example, see \cite{bonn}.

\newpage

\begin{figure}[h]
%Fig. 1
\centerline{\epsfig{file=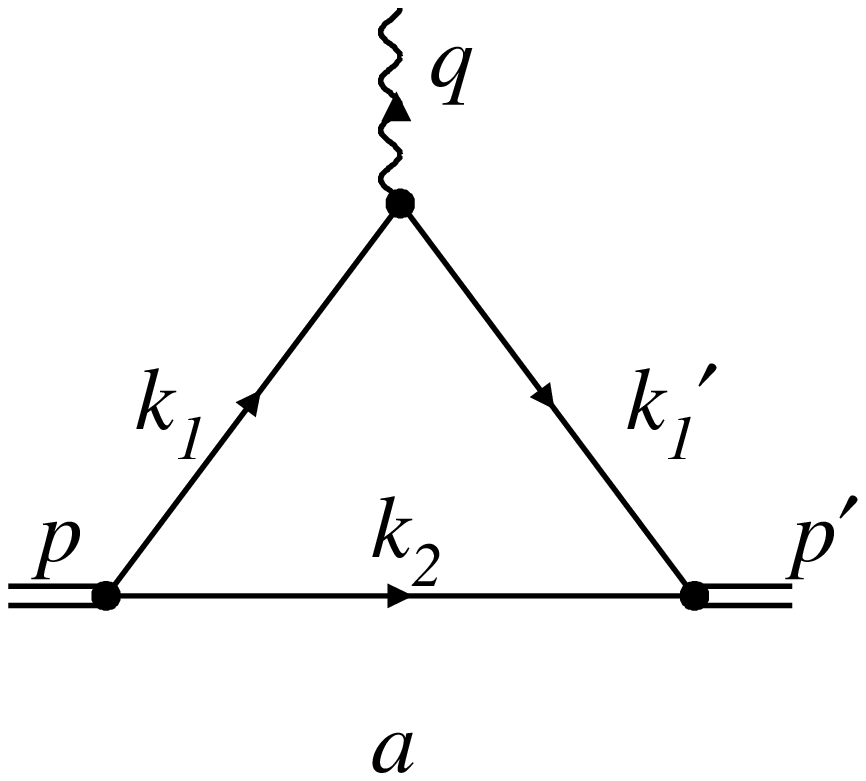,width=5cm}\hspace{0.5cm}
            \epsfig{file=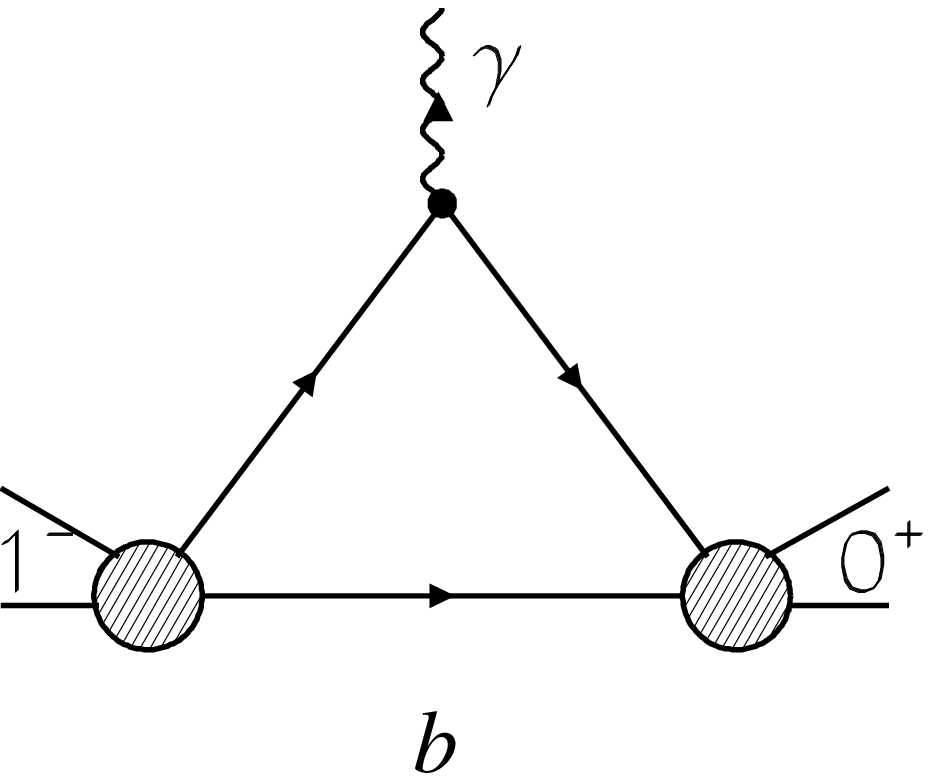,width=5cm}\hspace{0.5cm}
            \epsfig{file=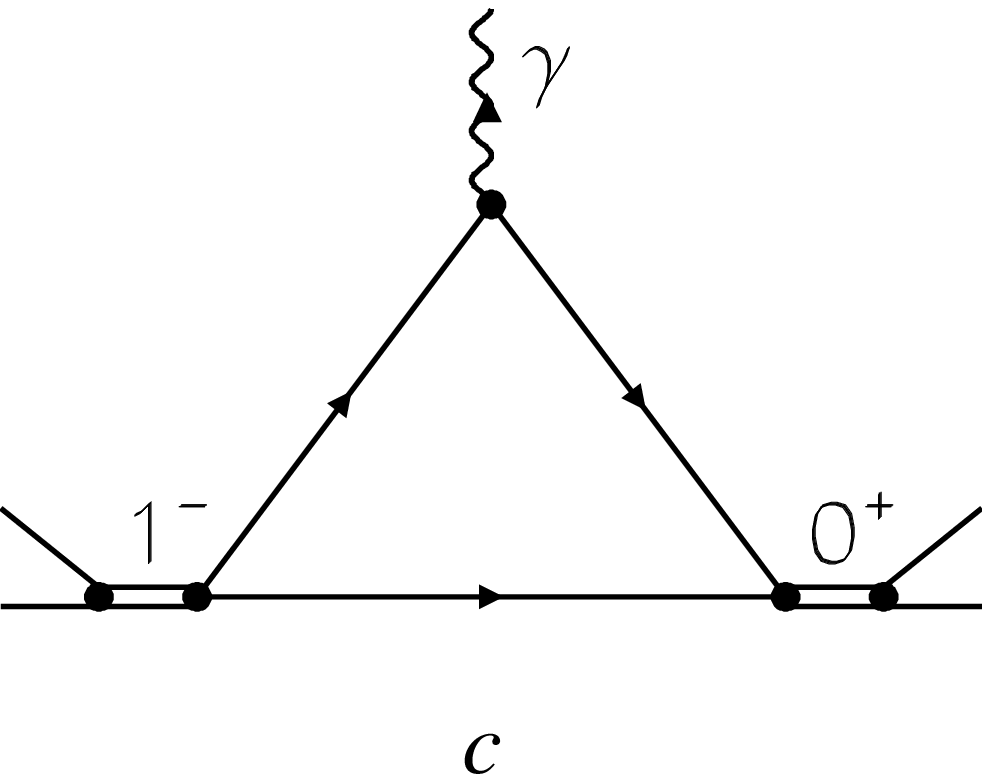,width=5cm}}
\caption{Transitions  $V\to\gamma S$ in the additive quark model.}
\end{figure}

\begin{figure}[t]
%Fig. 2
\centerline{\epsfig{file=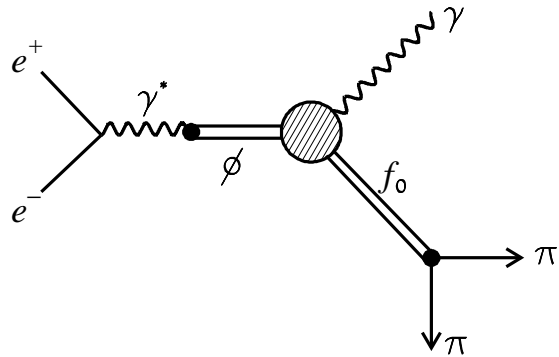,width=8cm}}
\caption{Process $e^+e^-\to \gamma\pi\pi$: residues in the
$e^+e^-$ and $\pi\pi$ channels determine the
$\phi \to \gamma f_0$ amplitude.}
\end{figure}

\begin{figure}[h]
%Fig. 3
\centerline{\epsfig{file=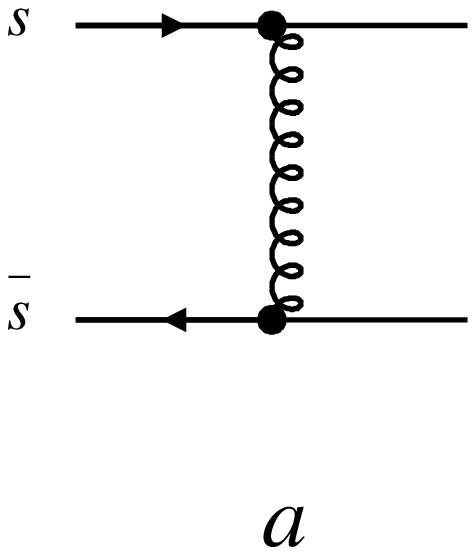,width=4cm}\hspace{0.4cm}
            \epsfig{file=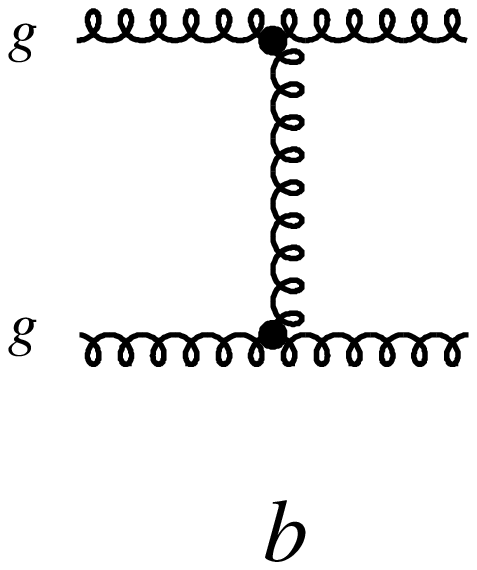,width=4cm}\hspace{0.4cm}
            \epsfig{file=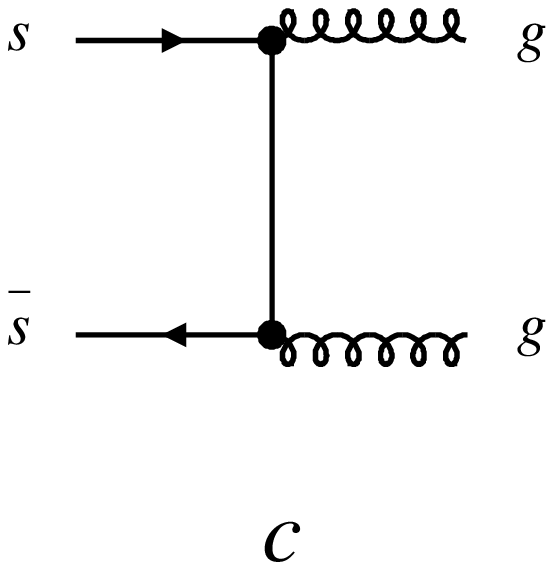,width=4cm}}
\caption{Examples of diagrams,  which contribute   to
the potentials
$U_{s\bar s\to s\bar s}(r)$, $ U_{s\bar s\to gg}({\bf r})$ and
$U_{gg\to gg}(r)$. }
\end{figure}

\begin{figure}[h]
%Fig. 4
\centerline{\epsfig{file=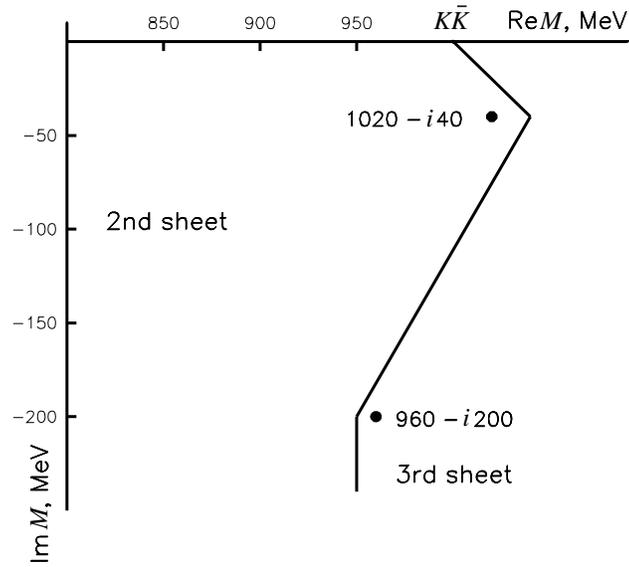,width=9cm}}
\caption{Complex-$M$ plane and location of the poles corresponding to
$f_0(980)$; the cut related to the $K\bar K$ threshold
is shown as broken line.}
\end{figure}

\begin{figure}[h]
%Fig. 5
\centerline{\epsfig{file=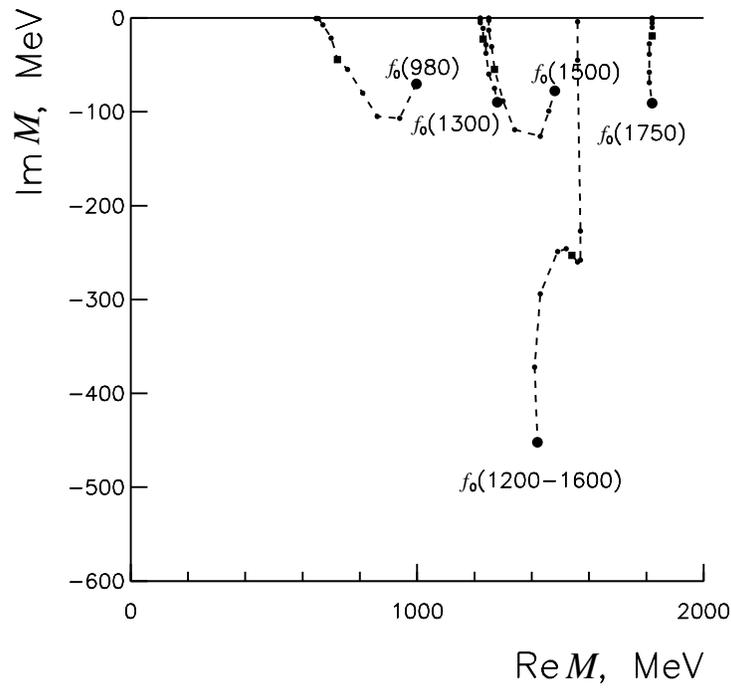,width=10cm}}
\caption{Complex $M$ plane: trajectories of poles
corresponding to the states $f_0(980)$, $f_0(1300)$, $f_0(1500)$,
$f_0(1750)$, $f_0(1200-1600)$ within a uniform onset of the decay
channels.}
\end{figure}

\begin{figure}[h]
%Fig. 6
\centerline{\epsfig{file=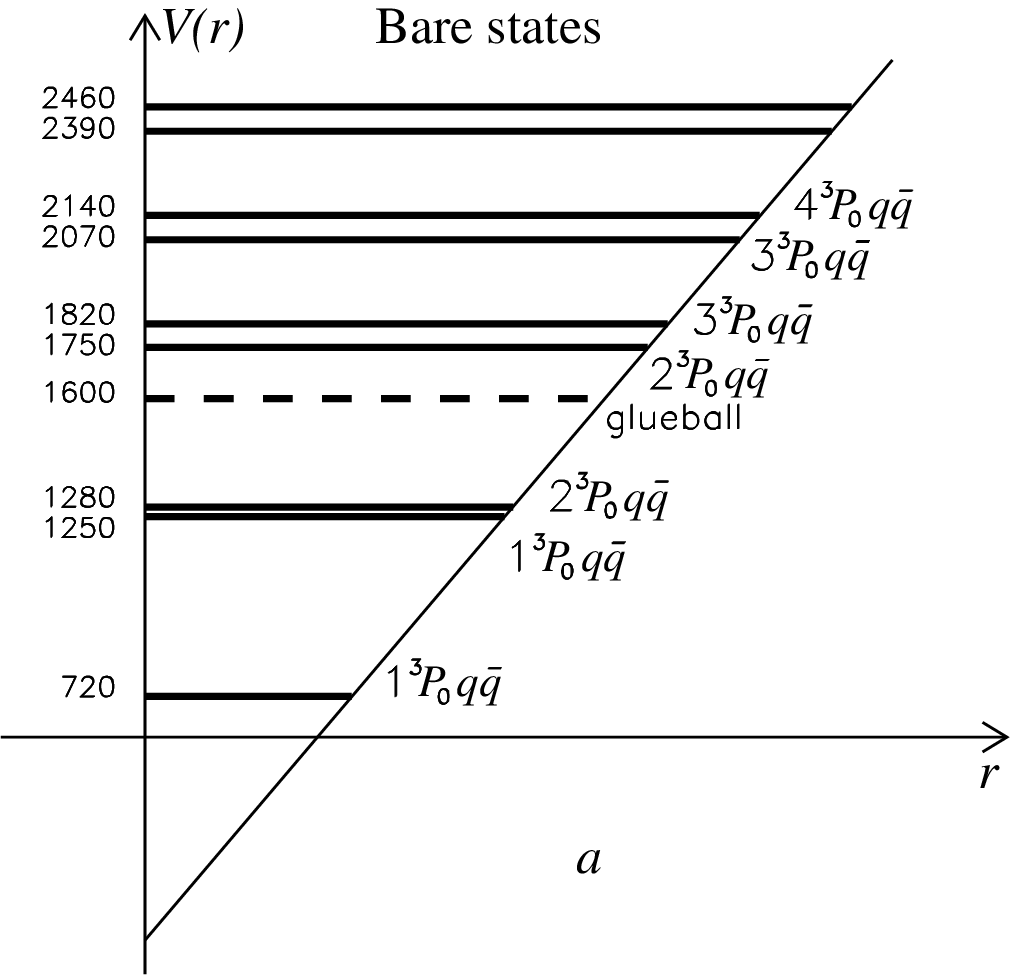,width=8cm}
            \epsfig{file=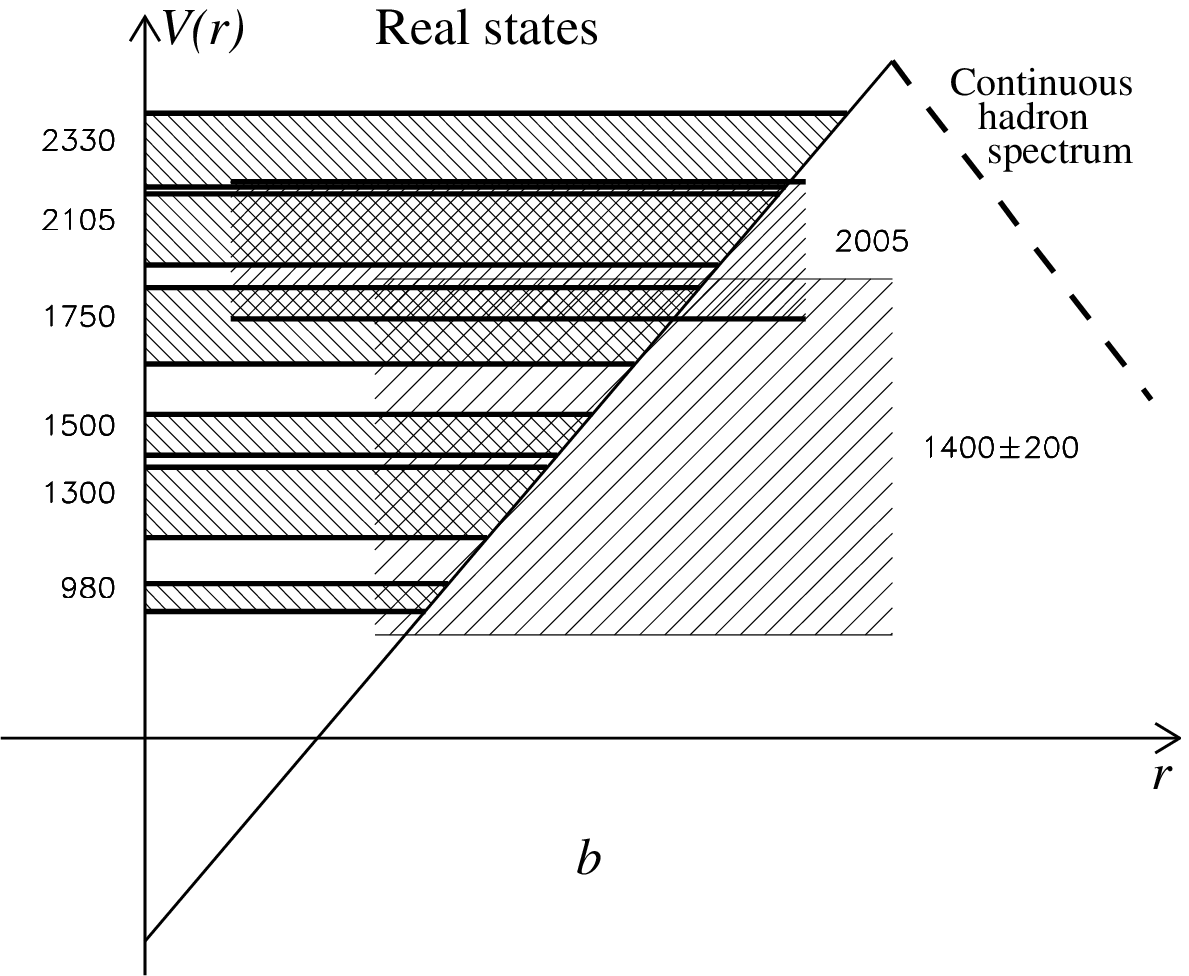,width=8cm}}
\caption{The $f_0$-levels in the potential well depending on
the onset of the decay channels: bare states (a)  and real resonances
(b).}
\end{figure}

\begin{figure}
%Fig. 7
\centerline{\epsfig{file=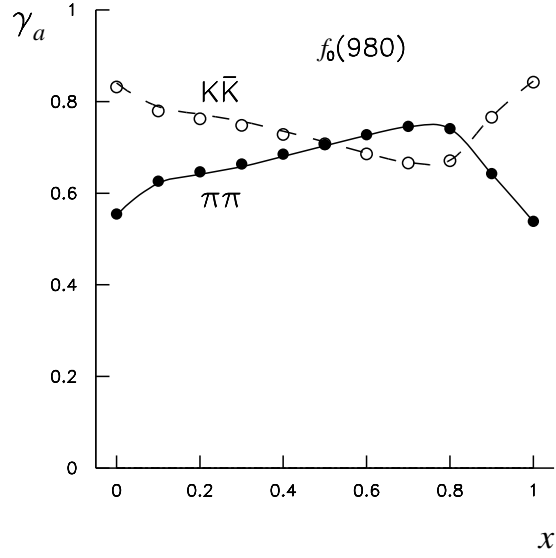,width=8cm}}
\caption{The evolution of normalized coupling constants
$\gamma_a =g_a/\sqrt{\sum_b g^2_b}$
at the onset of the decay channels for $f_0(980)$.}
\end{figure}

\newpage
\begin{figure}[h]
%Fig. 8
\centerline{\epsfig{file=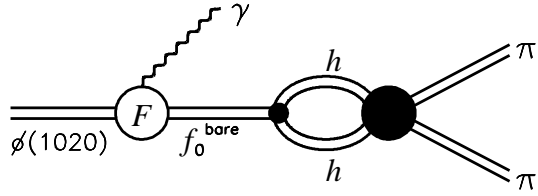,width=7cm}}
\caption{Diagram for the transition $\phi (1020)\to \gamma \pi\pi$
in the $K$-matrix representation, Eq. (46).}
\end{figure}

\begin{figure}[h]
%Fig. 9
\centerline{\epsfig{file=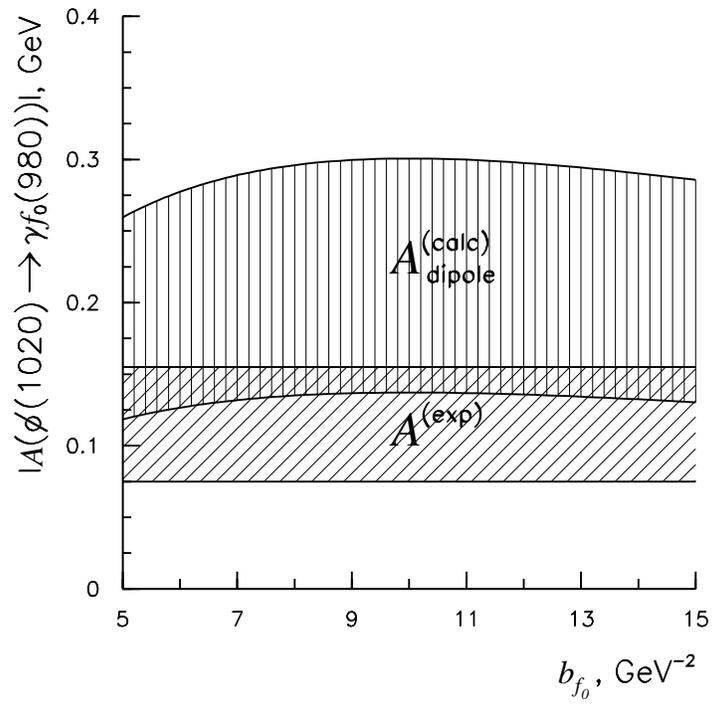,width=10cm}}
\caption{Amplitudes for the decay $\phi(1020)\to\gamma f_0(980)$:
the calculated ampitude $A^{(calc)}_{dipole}$ versus
the experimental one $A^{(exp)}$.}
\end{figure}

\begin{figure}[h]
%Fig. 10
\centerline{\epsfig{file=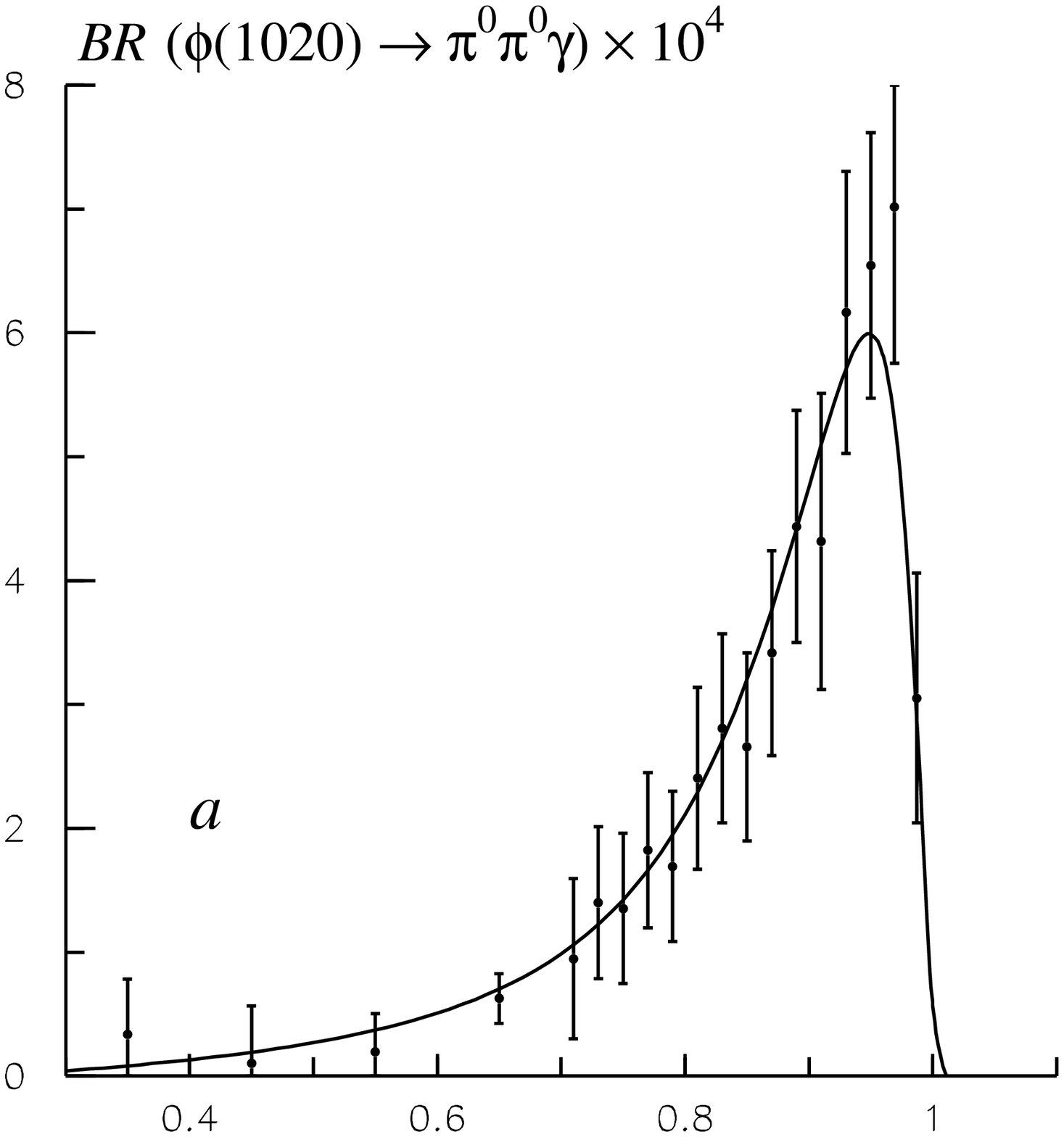,width=0.5\textwidth}}
\centerline{\epsfig{file=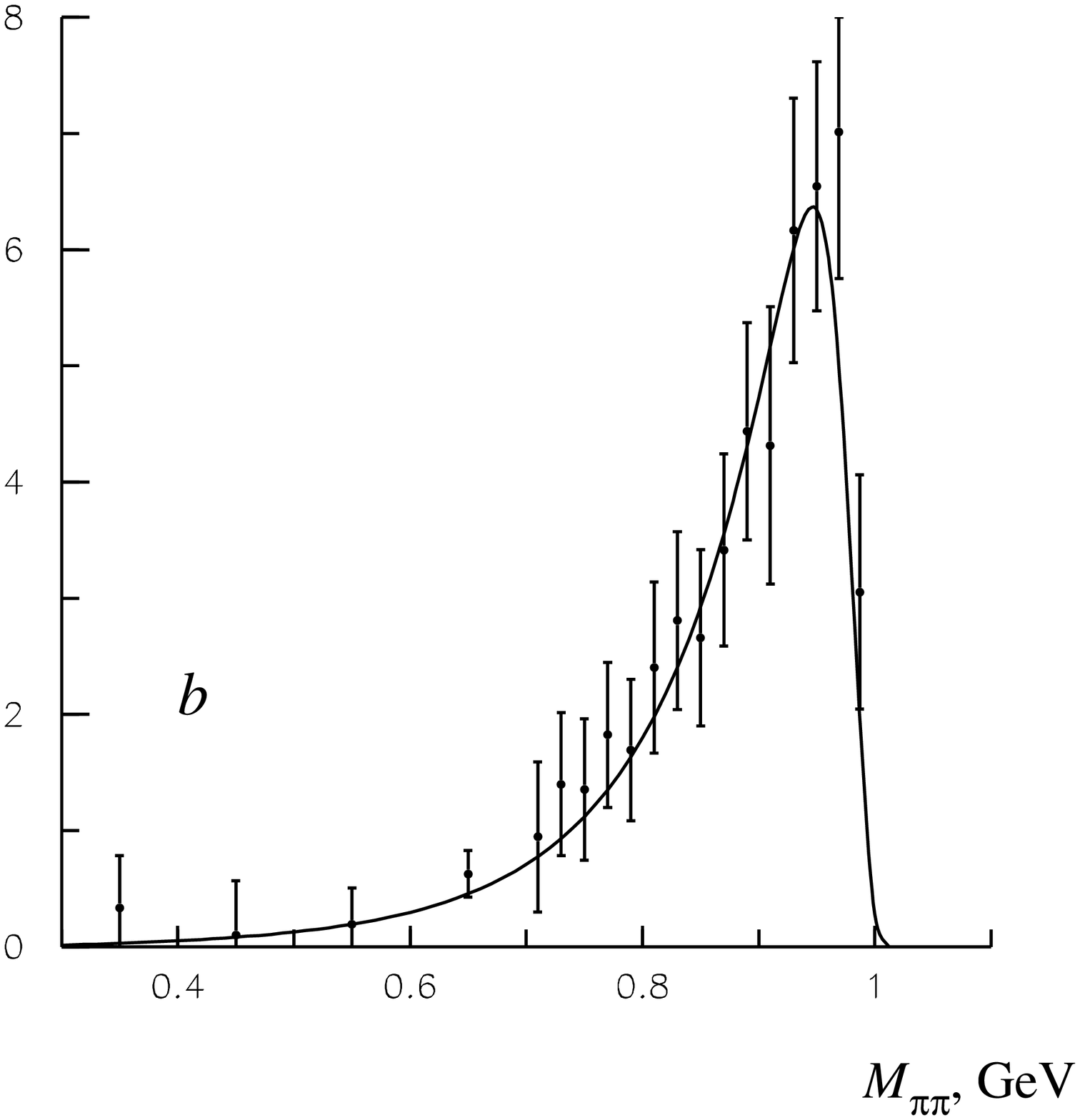,width=0.5\textwidth}}
\caption{The $\pi\pi$ spectrum of the reaction $\phi(1020)\to
\gamma\pi\pi$ calculated with the Flatte\'e formula (a) and Eq.
(68) (b).}
\end{figure}

\begin{figure}[h]
%Fig. 11
\centerline{  \epsfig{file=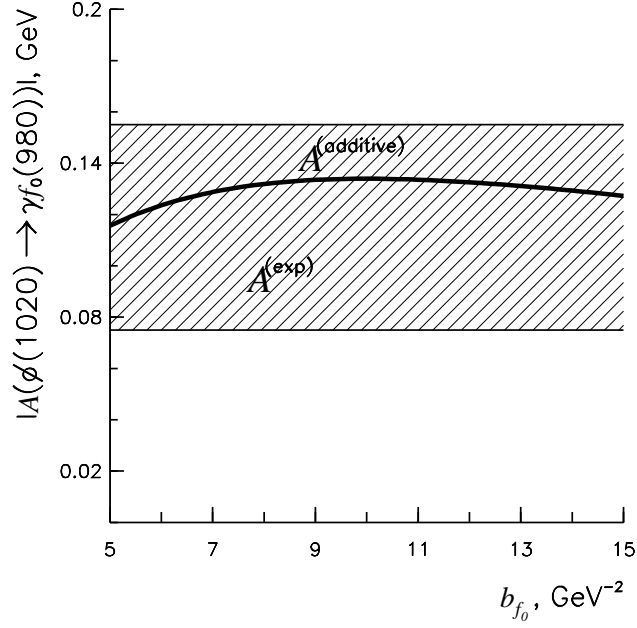,width=0.5\textwidth}}
\caption{ The additive quark model amplitude for
$\phi(1020)\to\gamma f_0(980)$, Eq. (72),  versus
$A^{(exp)}_{\phi(1020)\to\gamma f_0(980)}$ .  }
\end{figure}

\begin{figure}[h]
%Fig. 12
\centerline{\epsfig{file=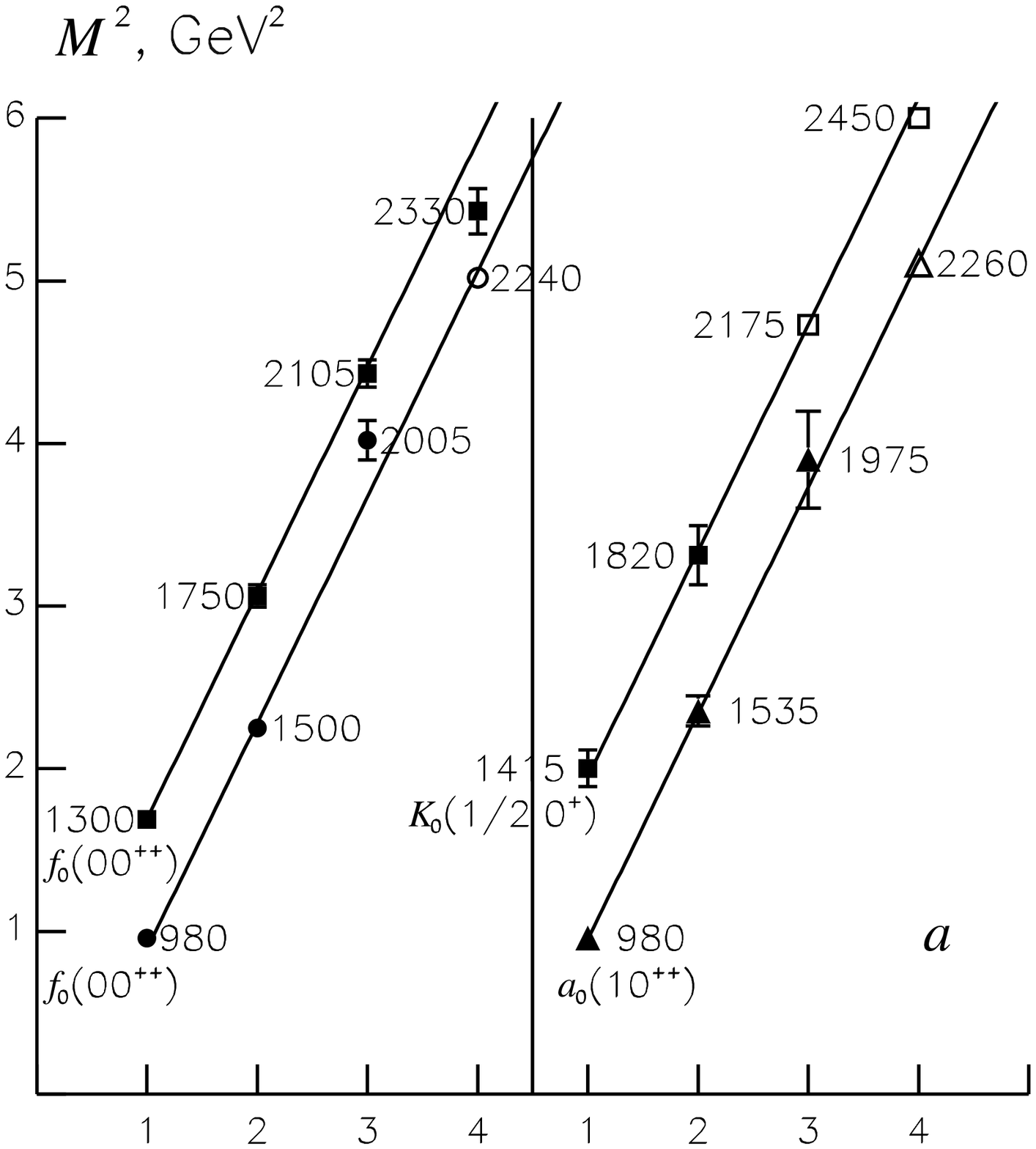,width=0.5\textwidth}
            \epsfig{file=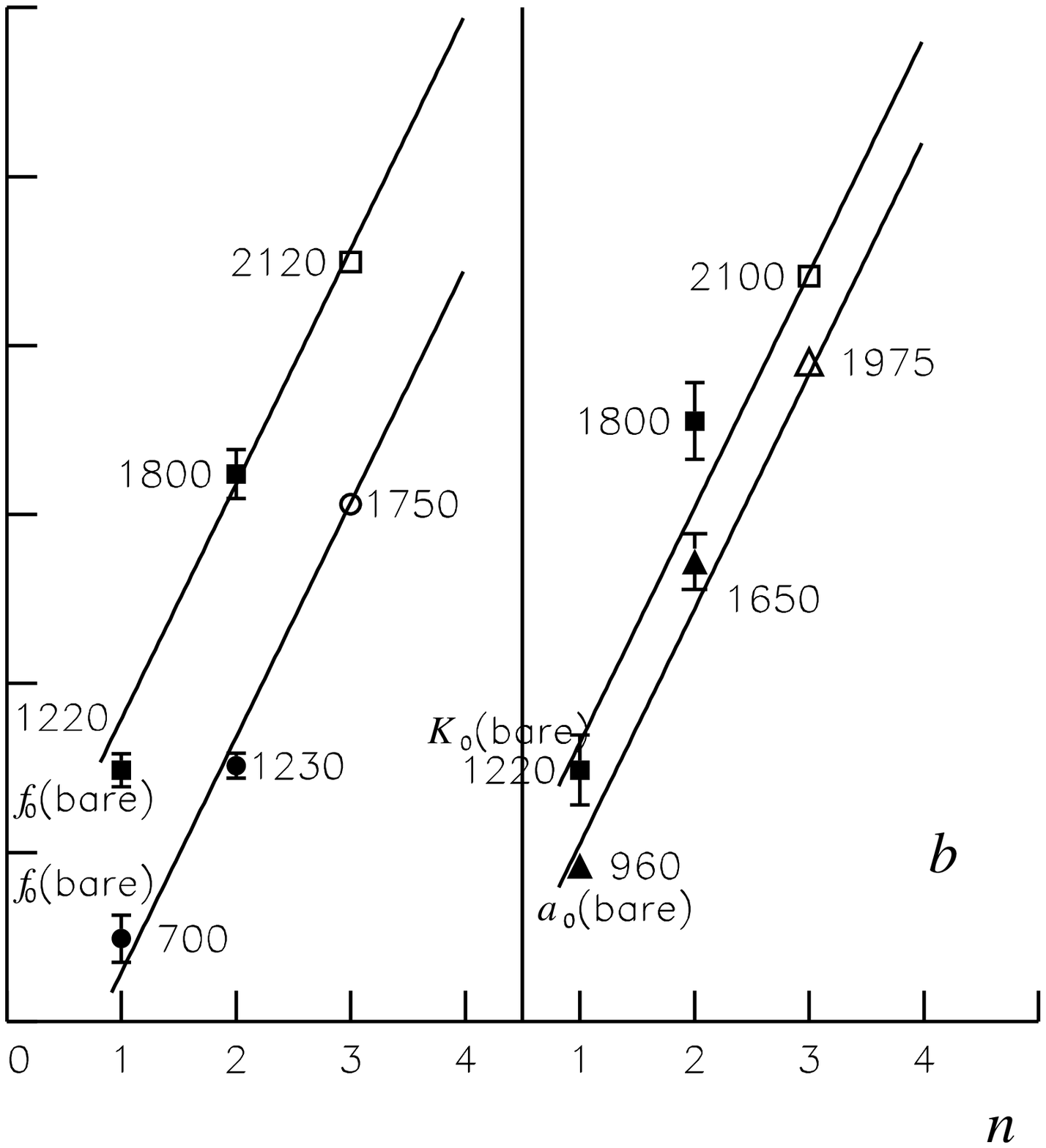,width=0.5\textwidth}}
\vspace{-0.5cm}
\caption{Linear trajectories on the
$(n,M^2)$ plane for bare states (a) and scalar resonances (b).}
\end{figure}


\begin{thebibliography}{99}

\bibitem{kmat} V. V. Anisovich and A. V. Sarantsev, Eur. Phys. J. A
{\bf 16}, 229 (2003).
\bibitem{prokoshkin}
 V. V. Anisovich, A. A. Kondashov, Yu. D. Prokoshkin,
S .A. Sadovsky, and A.V. Sa\-ran\-tsev, Yad.  Fiz. {\bf60}, 1489 (2000)
[Phys. At. Nucl. {\bf 60},  1410 (2000)].

\bibitem{km96}
V. V. Anisovich and A. V. Sarantsev, Phys. Lett. B {\bf 382}, 429
(1996);\\
V. V. Anisovich, Yu. D. Prokoshkin and A. V. Sarantsev, Phys.
Lett. B {\bf 389}, 388 (1996).

\bibitem{syst}
 A. V. Anisovich, V. V. Anisovich and A. V. Sarantsev,
Phys. Rev. D {\bf 62}:051502 (2000);
 V. V. Anisovich, hep-ph/0310165.



\bibitem{ufn} V. V. Anisovich, UFN {\bf 174}, 49 (2004);
hep-ph/0208123 v3.



\bibitem{jaffe} R. L. Jaffe,  Phys. Rev. D {\bf 15},
267 (1977).
\bibitem{isgur}J. Weinshtein and N. Isgur, Phys. Rev. D {\bf 41},
2236 (1990).

\bibitem{close-v}
F. E. Close $et al.$ Phys. Lett. B  {\bf 319}, 291 (1993).

\bibitem{8} N. Byers and R. MacClary, Phys. Rev. D {\bf 28}, 1692
(1983).
\bibitem{9} A. LeYaouanc, L. Oliver, O. Pene, and J.C. Raynal, Z. Phys.
{\bf C40}, 77 (1988).
\bibitem{siegert} A. J. F. Siegert, Phys. Rev. {\bf 52}, 787 (1937).


\bibitem{gauge} V. V. Anisovich and M. A. Matveev, hep-ph/0303119.

\bibitem{ital} S. Malvezzi,
 Talk given at {\it "Hadron-03", Aschaffensburg, Germany,
30 August-6 September 2003}.

\bibitem{ale-K} A. V. Anisovich and V. V. Anisovich, Phys. Lett. B
{\bf 467}, 289 (1999).
\bibitem{book} V. V. Anisovich, M. N. Kobrinsky, J. Nyiri and Yu. M.
Shabelski, {\it Quark Model and High Energy Collisions}, 2nd Ed.
(World Scientific, Singapore, 2004).




\bibitem{epja}
A. V. Anisovich, V. V. Anisovich, and V. A. Nikonov, Eur.
Phys. J. A {\bf12}, 103 (2001).
\bibitem{s}
A. V. Anisovich, V. V. Anisovich, V. N. Markov, and V. A.~Nikonov, Yad.
Fiz. {\bf65}, 523 (2002); [Phys. At. Nucl. {\bf65}, 497 (2002)].
\bibitem{nonrel} A. V. Anisovich, V. V. Anisovich, and V. A. Nikonov,
hep-ph/0305216.

\bibitem{lattice} G. S. Bali et al., Phys. Lett. B {\bf 309},
 378 (1993);
J. Sexton, A. Vaccarino, and D. Weingarten,
Phys. Rev. Lett. {\bf75} 4563  (1995);
C. J. Morningstar and M. Peardon, Phys. Rev. D {\bf56}, 4043 (1997).
\bibitem{PDG} D.  E. Groom  $et al$. (PDG), Eur. Phys. J. C {\bf 15}, 1
(2000).

\bibitem{content}
V. V. Anisovich, V. A. Nikonov, and A. V. Sarantsev,
Yad. Fiz. {\bf 65}, 1583 (2002); [Phys. At. Nucl. {\bf 65}, 1545
(2002)];
Yad. Fiz. {\bf 66}, 772 (2003); [Phys. At. Nucl.
{\bf 66}, 741 (2003)].


\bibitem{nov1}M. N. Achasov $et al.$ Phys. Lett. B {\bf 485}, 349
(2000).

\bibitem{nov2}R. R. Akhmetshin $et al.$ Phys. Lett. B {\bf 462},
380 (1999).
\bibitem{Flatte} S. M. Flatt\'e, Phys. Lett. B {\bf 63}, 224
(1976).

\bibitem{sarantsev}A. V. Sarantsev, private communication.
\bibitem{tuan}S. F. Tuan, hep-ph/0303248.
\bibitem{vanbeveren} E. van Beveren $et al.$ Mod. Phys. Lett. A {\bf
17}, 1673 (2002).
\bibitem{minkowski} W. Ochs and P. Minkowski, Nucl.
Phys. Proc. Suppl. {\bf 121}, 123 (2003).


\bibitem{glue1} G. Parisi and R. Petronzio, Phys. Lett. B {\bf 94}, 51
(1980);
 M. Consoli and J. H. Field, Phys. Rev. D {\bf 49}, 1293 (1994).

\bibitem{glue2} D. B. Leinweber $et al.$ Phys. Rev. D {\bf 58}, 031501
(1998).

\bibitem{dakhno} L. G. Dakhno and V. A. Nikonov, Eur. Phys. J.
{\bf A5}, 209 (1999).
\bibitem{penn-f}F. De Fazio and M. R. Pennington,
 Phys. Lett. B {\bf 521}, 15 (2001).
 \bibitem{deWitt}M. A. DeWitt, H. M. Choi and C. R. Ji,
Phys. Rev. D {\bf 68}, 054026 (2003).

\bibitem{LW}W. Lee and D. Weingarten,
Phys. Rev. D {\bf 61}, 014015 (1999).
\bibitem{CK} F. E. Close and A. Kirk, Eur. Phys. J. C {\bf 21}, 531
(2001).

\bibitem{close} F. E. Close, Int. J. Mod. Phys. A {\bf 17}, 3239 (2002).
\bibitem{achasov} N. N. Achasov, AIP Conf. Proc. {\bf 619}, 112 (2002).



\bibitem{gams}
 D. Alde  $et al.$ Zeit. Phys. C {\bf66}, 375 (1995);
Yu. D. Prokoshkin $et al.$ Physics -- Doklady {\bf 342},
473 (1995).

\bibitem{bnl}
J. Gunter $et al.$ (E852 Collaboration),
 Phys. Rev. D {\bf 64}, 07003 (2001).
\bibitem{chlap}
K. Ackerstaff $et al.$ Eur. Phys. J. C {\bf 4}, 19 (1998).

\bibitem{pp}  D. Barberis $et al.$ Phys. Lett.  B {\bf 453}, 325 (1999).

\bibitem{Ds}E. M. Aitola $et al.$ Phys. Rev. Lett. {\bf 86}, 765 (2001).
\bibitem{weDs} V. V. Anisovich, L. G. Dakhno and V. A. Nikonov,
hep-ph/0302137.
\bibitem{ochs} P. Minkowski and W. Ochs, Nucl. Phys. Proc. Suppl.
{\bf 121}, 119 (2003).

\bibitem{van}  F. Kleefeld $et al.$ Phys. Rev. D {\bf 66}, 034007
(2002).
\bibitem{gatto} A. Deandrea $et al.$ Phys. Lett. B {\bf 502}, 79
(2001).



\bibitem{pennington} M. Boglione and M. R. Pennington, Eur. Phys. J.
C {\bf 9}, 11 (1999).
\bibitem{tensor}
A. V. Anisovich, V. V. Anisovich, M. A. Matveev and V. A.~Nikonov,
Yad. Fiz.
{\bf66}, 946 (2003); [Phys. At. Nucl. {\bf 66}, 914 (2003)].

\bibitem{yu} Yu. S. Kalashnikova, private communication.

\bibitem{sz} Ya. B. Zeldovich and A. D. Sakharov, Yad. Fiz. {\bf 4}, 395
(1966); [Sov. J. Nucl. Phys. {\bf 4}, 283 (1967)].
\bibitem{glashow} A. de Rujula, H. Georgi, and S. L.~Glashow, Phys. Rev.
D {\bf 12}, 147 (1975).
\bibitem{bonn} R. Ricken, M. Koll, D. Merten,
B. C. Metsch, and H. R. Petry, Eur. Phys. J. A {\bf 9}, 221 (2000).



\end{thebibliography}
\end{document}